\documentclass[aps,prb,twocolumn,superscriptaddress,showpacs]{revtex4}
\usepackage{amssymb}
\usepackage{amsmath}
\usepackage{graphicx}

\begin{document}

\title{Continuous theory of ferroelectric states in ultrathin films with
real electrodes}
\author{A.M. Bratkovsky}
\affiliation{Hewlett-Packard Laboratories, 1501 Page Mill Road, MS 1123, Palo Alto,
California 94304}
\author{A.P. Levanyuk$^{1,}$}
\affiliation{Facultad de Ciencias, C-III, Universidad Autonoma de Madrid, University
Autonoma Madrid, Madrid 28049, Spain}
\date{December 28, 2007}

\begin{abstract}
According to a continuous medium theory, in very thin ferroelectric films
with real metallic electrodes (or dead layers near the electrodes)\ the
domain structure reduces to sinusoidal distribution of ferroelectric
polarization. Such a sinusoidal structure was considered in 1980s for
para-ferroelectric phase transition in a capacitor with dead layers near
electrode. We give a review of this theory and its further development for
the case of real metallic electrodes. The goal of the general theory is to
consistently interpret the experimental data in very thin films with real
metallic electrodes. This is illustrated on a recent experimental data for
5-30 nm BaTiO$_{3}$ films with SrRuO$_{3}$/SrTiO$_{3}$ electrodes. The
screening length by real metallic electrodes is very small ($<1${\AA }), but
it has a profound effect on ferroelectric properties and its phase behavior.
This general theory also allows to formulate the important open problems and
show paths towards their solution. In particular, this is a problem of
finding parameters of the system, which can sustain the ferroelectric memory
over a desired lifetime.
\end{abstract}

\maketitle
\tableofcontents

\section{Introduction}

Apparently, the continuous medium theory of domain structures in
ferroelectrics (FE) becomes even more important nowadays than in prior
years, when the studies concentrated on properties of the bulk samples. One
of the reasons is that the phenomenological domain theory, established
decades ago \cite{LL35,kittel,mitsui}, consistently captures the
electrostatic and strain effects playing crucial role in very thin high
quality FE films available now, and allows to gain an indispensable
qualitative insight into the phenomena at hand. One expects that the theory
should break down in atomically thin ferroelectrics, where the calculations
should be done from the first principles. This is formally true, yet the
theory applies to most systems of interest that are not just a few
monolayers thin, since the atomic length is so small. And the FE systems of
most interest to e.g. memory applications would not be that thin. In fact,
they should be thick enough to support a memory of practical interest, and
we reach this conclusion safely within the domain of applicability of the
continuum theory, as described below. The present review has a restricted
goal: it is aimed at explaining what the continuous medium theory tells us
about the properties of thin films with real metallic electrodes, especially
as far as domains and ferroelectric memory are concerned. We shall discuss
not only the theory but also some recent data, and we hope that as a result
of this discussion, the important role of the phenomenological theory as
well as necessary further developments of this theory, will become rather
evident

The principal point of the review is that some old theoretical results about
specific features of domain structures near a second order FE transition,
which were mainly of academic interest, gained practical importance now. Let
us recall that close to a second order transition the ferroelectric domain
structure cannot remain the same as it were far from the transition, where
domain walls thickness is smaller than the width of the domains. The reason
it that the domain wall thickness tends to infinity when the temperature
approaches the critical temperature $T_{c}$, $T\rightarrow T_{c}$, while the
period of the standard domain structure remains almost temperature
independent. As a result, some distance away from $T_{c}$ the width of the
domain walls becomes comparable with the width of the domains, i.e. the
spatial distribution of the polarization, which represents the domain
structure, becomes pretty smooth (\textquotedblleft sinusoidal"). One can
picture this change as elimination of higher Fourier harmonics of the
initially staggered distribution of the polarization across the domain
structure. The main harmonic is, obviously, the last to disappear since it
defines the main sluggish feature of the domain structure: its period.
Disappearance of the main harmonic is the phase transition into the
paraelectric phase, which occurs somewhere at temperature, $T<T_{c}.$ For
macroscopic samples with thickness $l\gg d_{at},$ where $d_{at}$ is the
characteristic interatomic distance, the elimination of the higher harmonics
is in fact quite sharp: all the above described transformation occurs within
a tiny, experimentally inaccessible temperature interval near $T=T_{c}.$ The
thinner the film the broader this interval, and when the film thickness
approaches the unit cell size, it may become quite important. Indeed, the
period of the domain structure is about $\sqrt{d_{at}l}$, Ref.~\cite{LL8}
(see also \cite{BL2000,BLPRB01,BLinh02} and our discussion below), while far
from $T_{c}$ the domain wall thickness can be roughly estimated as $d_{at}$
for the order-disorder transitions and $d_{at}\sqrt{T_{at}/T_{c}}$ for the
displacive ones, where $T_{at}\sim 10^{4}-10^{5}$K is the so called
characteristic \textquotedblleft atomic\textquotedblright\ temperature.
These estimates are fairly rough, for example in BaTiO$_{3},$ which is
usually considered a displacive ferroelectric, the $180-$degree domain walls
parallel to the direction (100) and the equivalent ones are actually very
narrow with the width of about $d_{at}$ at room temperature. However, the
estimates clearly indicate that in very thin films, with thicknesses
approaching $d_{at}$ and for displacive systems, and \textit{maybe } even in
thicker films not far from the phase transition, the domain width and the
domain wall thickness become comparable, so that the domain structure is
expected to be \textquotedblleft sinusoidal\textquotedblright .

It is worth mentioning that the basic physics of an effect of depolarizing
field and domains on the phase transition\ was well understood since the
pioneering work of K\"{a}nzig's group \cite{kanzig53} in 1950s, but the case
of the \textquotedblleft sinusoidal" domain walls\ was first considered by
Chensky and Tarasenko (ChT) some 30 years later\cite{chensky82}. Historical
aspects of the topic are described in more detail in Appendix A. ChT were
interested in the second order phase transitions in ferroelectric films in
both non-electroded and electroded samples with a dielectric
(\textquotedblleft dead") layer separating the electrodes and the film. In
the FE capacitors with metallic electrodes, the role of the
\textquotedblleft dead layers" is played by the metallic electrode
interfacial regions over the Thomas-Fermi screening length. Within the
continuous medium theory the mathematical analogy between the two cases is
practically exact (see below). Therefore, the model considered by ChT was,
in fact, very general. The results attracted little attention so far, since
the sinusoidal domains were expected to exist in a minute temperature
interval near phase transition in the systems experimentally studied at the
time. Now, however, the situation is quite different. In ferroelectric films
of atomic thickness, the sinusoidal regime has been experimentally found to
exist in a temperature interval of about $100$K near the transition\cite%
{streifer02}. The approach itself, new results, and the relevance of the
theory to the present experimental studies of nanoferroelectrics need to be
exposed to the ferroelectrics community. The theory becomes of practical
interest, and new problems need to be addressed. We discuss, in particular,
the problem of a (meta)stable ferroelectric memory. While most people are
currently concentrating on the problem of a \emph{critical thickness} for
ferroelectricity, which is shown to approach the atomic limit, the real
practical problem is that of a ferroelectric \emph{memory}, i.e. a
(meta)stable state with a net ferroelectric polarization in the film. These
are two \emph{different} problems, and we are trying to highlight this
important distinction.

Below, in Sec. II we present a simple overview of the ChT treatment of
paraelectric-ferroelectric transition in a film with the ideal metallic
electrodes and the dead layers. At the end of this Section, we demonstrate
similarity of the cases of real metallic electrodes and the dead layer
treated in\cite{chensky82}. In very thin strained (and, as a consequence,
uniaxial) films considered here, where a domain structure is close to the
sinusoidal one, we show the distribution of the polarization, which is
reminiscent of the one for the domains with staggered distribution of
polarization (the so-called Kittel structure\cite{kittel}.)

In Secs. III, IV we consider the properties of sinusoidal domain structures,
mainly their response to an external field. The main results of these
Sections present a further development of ChT theory. The algebra involved
is relatively simple in cases when the Landau-Ginzburg-Devonshire (LGD) free
energy contains the powers of polarization $P\ $up to $P^{4}$ terms only (in
the bulk or in the case of metallic electrodes this would correspond to a
second order transition far from a tricritical point), or just $P^{2}$ \emph{%
and} $P^{6}$ terms (describing a tricritical transition in the same
conditions), while in a more general case of a second order and/or the first
order transitions it is fairly involved. We limit ourselves here by the two
above cases, which illustrate the main results well enough to enable us to
discuss the available experimental data in Sec.V. Importantly, irrespective
of the conductivity type of electrodes, metallic or semiconducting, the
imperfect screening of bound charge of the FE films leads to their splitting
into domains\cite{BLcm06}. The results \cite{BLapl06} are used to discuss
the experiments by the Noh's team on BaTiO$_{3}$ films with SrRuO electrodes
on SrTiO$_{3}$ substrate \cite{KimL05,KimAPL05,nohNodl06} in Sec.V. There,
we conclude that their 5nm thick sample is well above the critical thickness
for ferroelectricity, but, even by very optimistic estimates, it is below
what can be called the critical thickness for the (meta)stable\emph{\
ferroelectric memory} with a reasonable lifetime (e.g. in excess of $10^{3})$%
\cite{BLmem07}. Unfortunately, at present the theory cannot give the
realistic number for this thickness. Throughout the review, we omit the
so-called additional boundary conditions (ABCs),\ which include parameters
of the surfaces/interfaces and are often taken into account together with
conventional electrostatic boundary conditions\cite{Kretschmer,BL05}. For
the experimental system that we discuss here, these additional boundary
conditions appear inessential, and more details about the additional
boundary conditions can be found in Appendix B.

\section{Loss of stability of the paraelectric phase}

The transition point is simultaneously the point of a stability loss of the
two phases for any second order phase transition. We shall consider second
order transitions from paraelectric phase in an electroded ferroelectric
film. This transition can be either into a single domain or a multidomain
state. To decide what choice is being realized (and at what conditions) the
best way is to consider the loss of stability of the phase whose state we
know: the paraelectric phase.

\subsection{The method}

\begin{figure}[h]
\begin{centering}
\includegraphics [width=6.5cm]{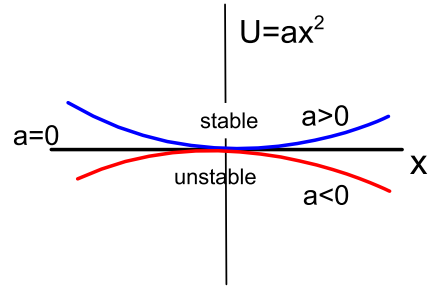}
\caption{
Schematic of the potential energy $U(x)=ax^2$ as a function of coordinate $x$
for various $a$: $a>0$, the state with $x=0$ is stable; $a=0$, point of stability loss of
the state with $x=0$; $a<0$, stability at $x=0$ is lost.
}\label{fig:Ushape}
\end{centering}
\vspace{0.05 cm}
\end{figure}

Let us explain how to determine when the loss of stability occurs by using
an elementary example. We show three forms of a potential energy $U\left(
x\right) =ax^{2}$ versus some coordinate $x$ in Fig.~\ref{fig:Ushape}.
Obviously, the state with $x=0$ is stable for $a>0$, for $a<0$ this state is
unstable, and $a=0$ (the horizontal line) corresponds to the point where the
stability is lost. The condition of equilibrium $dU/dx=0$ has only the
trivial solutions for $a\neq 0,$ but an infinite number of nontrivial
solutions for $a=0.$ For the potential energy of a more general form, $%
U\left( x\right) =ax^{2}+bx^{4}$ ($b>0$), we have the same point of the
stability loss, but $dU/dx=0$ has the trivial solution only for $a\geq 0.$
However, the \emph{linearized} form of $U$ near the origin $x=0$ is the same
as before. We have, therefore, the method of finding the conditions of
instability of a given state (or phase): one has to linearize the equations
whose solution represents the state in question and look for the conditions
when this system of equations has a nontrivial solution.


\begin{figure}[h]
\begin{centering}
\includegraphics [width=7cm]{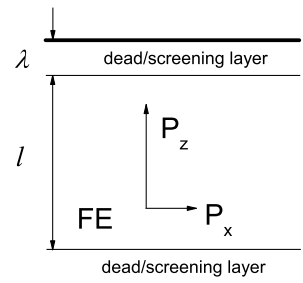}
\caption{ Schematic of the ferroelectric thin film with either dead
layers with thickness $d/2$ separating it from the electrodes, or
the real electrodes with Thomas-Fermi screening length
$\lambda$.}
\label{fig:FEfilmschema}
\end{centering}
\vspace{0.05 cm}
\end{figure}

We shall consider a uniaxial ferroelectric with the ferroelectric axis
perpendicular to the film plane, see Fig.~\ref{fig:FEfilmschema}.
This is the case, for example, of BaTiO$_{3}$ (BTO) films compressively
strained on SrRuO$_{3}$/SrTiO$_{3}$ (SRO/STO) substrate. The strained cubic
crystal behaves as a uniaxial ferroelectric with second- or weak first-order
phase transition. We restrict ourselves here to the case of a second order
ferroelectric transition in the bulk, i.e. the Landau-Ginzburg-Devonshire
(LGD) free energy density is assumed to have the form:
\begin{eqnarray}
F_{LGD}(P) &=&\frac{A}{2}P_{z}^{2}+\frac{B}{4}P_{z}^{4}+\ldots +\frac{1}{2}%
D_{ij}\left( \nabla _{\perp i}P_{z}\right) \left( \nabla _{\perp
j}P_{z}\right)  \notag \\
&&+\frac{1}{2}\eta \left( \partial _{z}P_{z}\right) ^{2}+\frac{1}{2}\kappa
P_{zb}^{2}  \label{LGD}
\end{eqnarray}%
where $P_{z}$ is the \emph{ferroelectric}{\normalsize \ }(switchable)\
component of polarization along the polar axis $z$, $P_{zb}$ is the%
{\normalsize \ \emph{nonferroelectric} }part of the polarization along the $%
z $-axis with{\normalsize \ }$\kappa $ directly related to the so-called
\emph{background} dielectric constant (see the Methodological Note below), $%
A=A^{\prime }\left( T-T_{c}\right) $ and $A^{\prime },B,$ $D_{ij},$ $\eta ,$
$\kappa =\mathrm{const}$, $\nabla _{\perp }=(\partial _{x},\partial _{y})$
the gradient operator in the plane $(x-y)$ of the film. To simplify the
formulas, we shall assume that the gradient coefficient is $D_{ij}=D\delta
_{ij}$, where $\delta _{ij}$ is the Kronecker symbol. Here, the Landau
coefficients are renormalized by the strain produced by the lattice misfit
with the substrate $u_{m}$ \cite{pertsev98}, see the Appendix D (for BTO on
STO\ $u_{m}=-2.2\%).$

Importantly, we do not include the striction terms that couple strain to $%
P^{2}$ terms in (\ref{LGD})\ because they renormalize the $BP_{z}^{4}$ term
and this does \emph{not} affect the loss of stability of symmetric phase in
a system like BTO on STO (see below). The point is that, as we said above,
finding the loss of stability boils down to solving a linear problem, while
this term is \emph{nonlinear} and does not affect the problem. The
statements to the contrary by Pertsev and Kohlstedt in Ref.\cite{perkohl07},
that the elastic coupling defines the point of the stability loss and the
critical thickness of the FE\ films, are qualitatively incorrect, as pointed
out in Ref.~\cite{BLcomPer07}. The above form of $F_{LGD\ }$ yields the
equation of state for the ferroelectric via standard relation:
\begin{equation}
E_{fz}=\frac{\delta F_{LGD}}{\delta P_{z}}=AP_{z}+BP_{z}^{3}-D\nabla _{\perp
}^{2}P_{z}-\eta \partial _{z}^{2}P_{z},  \label{eqstate1a}
\end{equation}%
\begin{equation}
E_{fz}=\kappa P_{zb},
\end{equation}%
which means that the local electric field is in one-to-one correspondence
with the local ferro- and nonferroelectric components of the polarization.

\subsection{Loss of stability with respect to homogeneous polarization}

Since we do not know what type of solutions corresponds to the horizontal
line in Fig.~\ref{fig:Ushape}, we should check all the possibilities. We
start by trying the simplest one, a homogeneous change of polarization. This
situation is described by the linearized equation of state for a \emph{%
homogeneous} polarization $P_{zh}$ (and the nonferroelectric part $\kappa
P_{zb}):$%
\begin{equation}
E_{fz}=AP_{zh}=\kappa P_{zb}  \label{eqstate1c}
\end{equation}%
where $E_{fz}$ is the homogeneous field in the ferroelectric. The dead layer
is a linear dielectric, which is described by an equation analogous to (\ref%
{eqstate1c}), hence it is not necessary to consider explicitly the
polarization of the dead layer, and we shall describe it by its dielectric
constant $\epsilon _{e}$. The boundary condition at the interface of the
ferroelectric (continuity of the normal component of the displacement
vector) then reads
\begin{equation}
E_{fz}+4\pi \left( P_{zh}+P_{zb}\right) =\epsilon _{e}E_{dz},  \label{eq:bcz}
\end{equation}%
where $E_{dz}$ is the field in the dead layer. We can rewrite the above
equation as%
\begin{equation}
\epsilon _{b}E_{fz}+4\pi P_{zh}=\epsilon _{e}E_{dz},  \label{bcz1}
\end{equation}%
where $\epsilon _{b}$ is the "background dielectric constant", $\epsilon
_{b}=1+4\pi /\kappa $. There is also a condition of a zero applied bias
(short circuiting, $U=0$):
\begin{equation}
E_{fz}l+E_{dz}d=0.  \label{short}
\end{equation}%
We find from the above the depolarizing field in the FE film:
\begin{eqnarray}
E_{fz} &=&-\frac{4\pi d}{\epsilon _{e}l+\epsilon _{b}d}P_{zh}\approx -\frac{%
4\pi dP_{zh}}{\epsilon _{e}l},  \label{dep} \\
E_{dz} &=&\frac{4\pi l}{\epsilon _{e}l+\epsilon _{b}d}P_{zh}\approx \frac{%
4\pi P_{zh}}{\epsilon _{e}},  \label{eq:Ed}
\end{eqnarray}%
and we see that the field in the dead layer is much stronger than the field
in the FE film, $\left\vert E_{dz}\right\vert \gg \left\vert
E_{fz}\right\vert $ because is does not contain the small parameter $d/l\ll
1 $, and it is realistic to assume $\epsilon _{b}\sim \epsilon _{e}$.
Importantly, the field is concentrated in the dead layers, yet since the
layers are very thin, its contribution to the total free energy appears to
be negligible (see below). The system of homogeneous equations~(\ref%
{eqstate1c}),(\ref{dep}) for $E_{fz}$\ and $P_{0}$\ has nontrivial solutions
only when
\begin{equation}
A=-\frac{4\pi d}{\epsilon _{e}l+\epsilon _{b}d}\approx -\frac{4\pi d}{%
\epsilon _{e}l}.  \label{eq:Ah}
\end{equation}%
This would define the temperature of the stability loss of the paraelectric
phase with regards to a \emph{homogeneous} polarization.{\normalsize \ }It
would happen at a lower temperature than the loss of stability of the
paraelectric phase in a very thick slab ($l\rightarrow \infty $) or in
absence of a dead layer and ideal metallic electrodes ($A=0,$\ $T=T_{c}$).
Recall that absence of nontrivial solutions at values of $A$\ which do not
satisfy Eq.(\ref{eq:Ah}) means either stability of the solution $P=0$\ (the
paraelectric phase) or instability of this solution (see Fig.1). and,
therefore, stability of another homogeneous solution with $P\neq 0,$\ which
has to be found from the full, nonlinearized, equation of state. Let us
emphasize that this is not necessarily the absolutely stable state but might
be a metastable state because we did not consider competing inhomogeneous
states which might have a lower energy. Also the temperature defined by Eq.(%
\ref{eq:Ah}) is $not$, in general, the temperature at which the paraelectric
phase ceases to exist: one has to consider a loss of stability with respect
to an \emph{inhomogeneous} polarization as well, in order to determine which
one sets in first.

Now, consider the case of\emph{\ a finite bias} voltage $U$. \ In this case,
the equilibrium polarization in the \emph{paraelectric} phase is non-zero,
and so are the electric fields in the ferroelectric and the dead layer. One
finds
\begin{equation*}
lE_{fz}+dE_{dz}=U
\end{equation*}%
and, instead of Eq.~(\ref{dep}), we obtain the field in the FE film:
\begin{eqnarray}
E_{fz} &=&E_{0}-\frac{4\pi d}{\epsilon _{e}l+\epsilon _{b}d}P_{zh},
\label{extdep} \\
E_{0} &=&\frac{U}{l+\epsilon _{b}d/\epsilon _{e}}\approx \frac{U}{l},
\label{eq:E0}
\end{eqnarray}%
where $E_{0}$ is the \emph{external field}. The expression for $E_{fz}$ (\ref%
{extdep}) is usually said to mean that the field in the FE film is equal to
a sum of the external $E_{0}$ and the \emph{depolarizing} fields.
Substituting Eq.~(\ref{extdep}) into Eq.~(\ref{eqstate1a}), we obtain
\begin{equation}
\left( A+\frac{4\pi d}{\epsilon _{e}l+\epsilon _{b}d}\right)
P_{0}+BP_{0}^{3}=E_{0},  \label{Pe}
\end{equation}%
where $P_{0}$ stands for the equilibrium homogeneous polarization.

Let us check if this equilibrium state can lose its stability with respect
to a homogeneous fluctuation of polarization. In textbooks on
ferroelectricity, it is shown that when $d=0$ (an ideal metallic electrodes)
there is no loss of stability, and the second order transition gets smeared
out by the external field. Lets check that the homogeneous transition would
also be smeared out in the case of the dead layers. Checking for the
stability loss with respect to a homogeneous polarization, we linearize the
equation of state (\ref{eqstate1a}) near $P=P_{0}$, $E_{fz0}$, where $%
E_{fz0} $\ is the electric field in the ferroelectric in the state with $%
P=P_{0}$, and obtain
\begin{equation}
\left( A+3BP_{0}^{2}\right) P^{\prime }=E_{fz}^{\prime },
\end{equation}%
where $P^{\prime }=P-P_{0}$ and $E_{fz}^{\prime }=E_{fz}-E_{fz0}.$ We obtain
the same equations for the fluctuation of the polarization $P^{\prime }$ and
the field $E_{fz}^{\prime }$ as for the short-circuited case with the only
difference that $A$ should be replaced by $\overline{A}=A+3BP_{0}^{2}.$ We
have already found [cf. Eq.~(\ref{eq:Ah})] that the loss of stability with
respect to a homogeneous polarization should take place if
\begin{equation}
\overline{A}=A+3BP_{0}^{2}=-\frac{4\pi d}{\epsilon _{e}l+\epsilon _{b}d}.
\label{eq:noloss}
\end{equation}%
This is \emph{impossible}, however, since from Eq.~(\ref{Pe}):
\begin{equation}
\overline{A}+\frac{4\pi d}{\epsilon _{e}l+\epsilon _{b}d}=A+\frac{4\pi d}{%
\epsilon _{e}l+\epsilon _{b}d}+3BP_{0}^{2}=\left( \frac{dP_{0}}{dE_{0}}%
\right) ^{-1},
\end{equation}%
is always \emph{positive}. Indeed, $dP_{0}/dE_{0}$ can be considered as the
differential susceptibility of a ferroelectric in an external electric field
and with the ideal metallic electrodes, but with the coefficient before the $%
P^{2}$ term equal to $A+4\pi d/\left( \epsilon _{e}l+\epsilon _{b}d\right) $%
. We have already mentioned that in the case of the ideal metallic
electrodes there is no loss of stability and the phase transition is
smeared, i.e. $dP_{0}/dE_{0}$ is always positive under a finite bias
voltage. We have proven that in a ferroelectric with the dead layers (or
\emph{real} metallic electrodes) the phase transition with respect to the
\emph{homogeneous} polarization is smeared out by an external electric
field. As was mentioned above, this does \emph{not} mean that there is no
phase transition. Indeed, there is a possibility that in an external field
the film may be losing stability with respect to inhomogeneous polarization.
If this takes place, a second order (unsmeared) phase transition occurs \
This is why we shall study a loss of stability with respect to the
inhomogeneous polarization in a general case of applied external bias
voltage.

\subsection{Loss of stability with respect to inhomogeneous polarization}

Following the same general procedure, we should be looking for a first
occurrence of a nontrivial solution to a linearized equation of state that
allows for inhomogeneous polarization:
\begin{equation}
E_{z}=\overline{A}P_{z}-D\nabla _{\perp }^{2}P_{z}-\eta \partial
_{z}^{2}P_{z}.  \label{eqstate1b}
\end{equation}%
Since an inhomogeneous polarization $\widetilde{P}_{z}$ produces an electric
field both across the film ($z-$axis) and in perpendicular directions, we
should add an equation of state for the two other polarization components.
We have:
\begin{equation}
\boldsymbol{P}_{\perp }=\frac{\epsilon _{\perp }-1}{4\pi }\boldsymbol{E}%
_{\perp },
\end{equation}%
where $\boldsymbol{E}_{\perp }$ is the in-plane electric field, $\epsilon
_{\perp }$ the in-plane dielectric constant. The solution of (\ref{eqstate1b}%
) should also obey the equations of electrostatics:
\begin{equation}
\boldsymbol{\nabla }\times \boldsymbol{E}=0,  \label{el1}
\end{equation}%
\begin{equation}
\boldsymbol{\nabla \cdot }\left( \epsilon _{b}\boldsymbol{E}+4\pi
\boldsymbol{P}\right) =0.  \label{el2}
\end{equation}%
Now, we should get inventive. As far as a reasonable solution of equations (%
\ref{eqstate1b}), (\ref{el1}), (\ref{el2}) is concerned, they have a vast
set (continuous infinity) of types of solutions and we have to single out
only one type of solutions from them that gets realized. We shall apply a
physical common sense to restrict the types of the solutions considered.
First, the prospective solutions should be periodic in the film plane, since
we deal with a large area (practically infinite) film. Any periodic function
can be presented as a sum of the sinusoidal functions. Consider the
structure that is periodic in the $x$-direction. Then,
\begin{equation}
\widetilde{P}_{z}(x,z)=\sum_{n}\widetilde{P}_{n}\left( z\right) \cos k_{n}x,
\label{periodic}
\end{equation}%
where $k_{n}=nk_{1},$ $\widetilde{P}_{n}\left( z\right) $ are the unknown
functions and $k_{1}$ is the unknown wavenumber. We have to perform the
stability check for arbitrary $k_{1}$. The solution given by Eq.~(\ref%
{periodic}) is very complicated, and in reality one does not need to keep
all harmonics. It is sufficient to consider the critical harmonic with a
one-dimensional periodicity
\begin{equation}
\widetilde{P}_{z}=af\left( z\right) \cos kx,  \label{Ansatz}
\end{equation}%
where $a$ and $k$ are the unknown numbers, $f\left( z\right) $ the unknown
function. It adds nothing new if we consider, e.g., $\cos k_{1}x\cos k_{2}y$%
, because the products of trigonometric functions are the linear
combinations of trigonometric functions of other (sum and difference)\
arguments. Looking among the forms given by Eq.~(\ref{Ansatz}), we have
already tremendously simplified the problem: for every $k$ we will have a
system of ordinary differential equations, instead of a system of partial
differential equations. Further, we can identify $f\left( z\right) $ too.
First of all, this function should be as smooth as possible. Indeed, the
bound charge associated with the inhomogeneous ferroelectric polarization $%
\rho _{b}=-\boldsymbol{\nabla \cdot P}=a\left( df/dz\right) \cos kx$ should
be as small as possible to reduce the electric field energy. It is
unreasonable to have $f\left( z\right) =\mathrm{const}$, the bound charge at
the ferroelectric-dead layer interface will be proportional to this
constant. More reasonable would be to make it nearly constant deep inside
the ferroelectric yet smaller at the interface. Taking into account that it
should be a solution of a differential equation with the constant
coefficients, we are left with an ansatz:
\begin{equation}
\widetilde{P}_{z}=a\cos kx\cos qz,  \label{Ansatz1}
\end{equation}%
where $q$ is another unknown parameter, which we expect to be much smaller
than $k.$ A critical reader could study other options. It may seem formally
possible to try $\cosh qz,$ for example, but this is a bad choice that fails
to satisfy the boundary conditions (see below), which are possible to
satisfy with Eq.~(\ref{Ansatz1}).

Using Eq.~(\ref{eqstate1b}), one obtains for the ansatz (\ref{Ansatz1}):
\begin{equation}
\widetilde{E}_{fz}\left( x,z\right) =\left( \overline{A}+Dk^{2}+\eta
q^{2}\right) a\cos kx\cos qz.  \label{Ezf}
\end{equation}%
The electrostatics equation (\ref{el1})$\ $yields
\begin{equation}
\partial _{z}\widetilde{E}_{fx}=\partial _{x}\widetilde{E}_{fz},
\label{eq:curlE}
\end{equation}%
and one finds that
\begin{equation}
\widetilde{E}_{fx}\left( x,z\right) =-\frac{k}{q}\left( \overline{A}%
+Dk^{2}+\eta q^{2}\right) a\sin kx\sin qz.  \label{Exf}
\end{equation}%
Thus, we have found the electric field in the ferroelectric assuming\emph{\ }%
that we have a \textquotedblleft polarization wave\textquotedblright . This
assumption is consistent if this field satisfies the second electrostatic
equation (\ref{el2}), or
\begin{equation}
\epsilon _{b}\partial _{z}\widetilde{E}_{fz}+\epsilon _{\perp }\partial _{x}%
\widetilde{E}_{fx}+4\pi \partial _{z}\widetilde{P}_{z}=0.  \label{eq:poisPz}
\end{equation}%
We can easily get rid of $\widetilde{E}_{fx}$ in the above equation by
differentiating it with respect to $z$ and using Eq.(\ref{eq:curlE}) to
obtain
\begin{equation}
\left( \epsilon _{b}\partial _{z}^{2}+\epsilon _{\perp }\partial
_{x}^{2}\right) \widetilde{E}_{fz}+4\pi \partial _{z}^{2}\widetilde{P}_{z}=0.
\label{eq:Etz}
\end{equation}%
Substituting here the ansatz (\ref{Ezf}), we find the value of $\overline{A}%
, $ at which the non-trivial solutions are possible as a function of\ the
parameters $k$ and $q$:
\begin{equation}
\overline{A}=-\frac{4\pi q^{2}}{\epsilon _{b}q^{2}+\epsilon _{\perp }k^{2}}%
-Dk^{2}-\eta q^{2}.  \label{eq:Akq}
\end{equation}

It is still premature to look for a maximum of this function (a point, where
the instability sets in first), because we have also (i)\ to find the fields
from the electrostatic equations in the dead layer and (ii)\ to satisfy the
boundary conditions. Given that the field in the dead layer has to have its $%
z$ component depending on $x$ as $\cos kx$, we put $\widetilde{E}_{dz}=\zeta
\left( z\right) \cos kx$ and, therefore, $\widetilde{E}_{dx}=\psi \left(
z\right) \sin kx$, where $\psi ^{\prime }\left( z\right) =-k\zeta \left(
z\right) $. From $\boldsymbol{\nabla \cdot E}=0,$ one has $\zeta ^{\prime
}+k\psi =0$ or $\psi ^{\prime \prime }-k^{2}\psi =0$. Since the transversal
component of the electric field is zero at an ideal metallic surface, we
find that $\psi \left( z\right) =\Psi \sinh k\left( z-l/2-d/2\right) $ and
\begin{eqnarray}
\widetilde{E}_{dx} &=&\Psi \sinh k\left( z-l/2-d/2\right) \sin kx,
\label{eq:Edx} \\
\widetilde{E}_{dz} &=&-\Psi \cosh k\left( z-l/2-d/2\right) \cos kx,
\label{eq:Edz}
\end{eqnarray}%
for $z>l/2.$ The two electrostatic boundary conditions, (i) the continuity
of the in-plane electric field component and (ii)\ the continuity of the
displacement, Eq.~(\ref{eq:bcz}),\ at the ferroelectric-dead layer interface
($z=l/2$) read
\begin{equation}
a\frac{k}{q}\left( \overline{A}+Dk^{2}+\eta q^{2}\right) \sin \frac{ql}{2}%
=\Psi \sinh \frac{kd}{2},  \label{bc1}
\end{equation}%
\begin{equation}
a\left( \overline{A}+4\pi /\epsilon _{b}+Dk^{2}+\eta q^{2}\right) \cos \frac{%
ql}{2}=-\frac{\epsilon _{e}}{\epsilon _{b}}\Psi \cosh \frac{kd}{2}.
\label{bc2}
\end{equation}%
Excluding $\overline{A}$ from equations (\ref{bc1}),(\ref{bc2}),\ with the
help of Eq{\normalsize .}(\ref{eq:Akq}), one finds that this system has a
non-trivial solutions for $a$ and $\Psi $ if
\begin{equation}
\frac{q}{\epsilon _{\perp }}\tan \frac{ql}{2}=\frac{k}{\epsilon _{e}}\tanh
\frac{kd}{2}.  \label{kq1}
\end{equation}

Since we are interested in the case of thin dead layers, we suppose that $%
kd\ll 1$. Then, Eq.~(\ref{kq1}) simplifies to
\begin{equation}
q\tan \frac{ql}{2}=\frac{\epsilon _{\perp }k^{2}d}{2\epsilon _{e}}.
\label{kq2}
\end{equation}%
This allows one to exclude $k$ from (\ref{kq2})\ and (\ref{eq:Akq})\ and
reduce, for the case of thin dead layer, i.e. for $l\gg d\epsilon
_{b}/\epsilon _{e},$ $d\eta \epsilon _{\perp }/\left( D\epsilon _{e}\right)
, $ to:$.$%
\begin{eqnarray}
-\overline{A} &=&\frac{4\pi d}{\epsilon _{e}l}t(y),\qquad t(y)=y\left( \cot
y+r\tan y\right) ,  \label{eq:Atw} \\
r &=&\frac{\epsilon _{e}^{2}D}{\pi \epsilon _{\perp }d^{2}},\qquad y=ql/2.
\end{eqnarray}%
Now, we can find at what critical thickness of the dead layer the first
nontrivial solution appears with $q\neq 0.$ Consider first the case $%
y=ql/2\ll 1,$ where the function $t(y)\ $can be expanded as
\begin{equation}
t\approx 1+\left( r-\frac{1}{3}\right) y^{2}+\left( \frac{r}{3}-\frac{1}{45}%
\right) y^{4}.
\end{equation}%
There appears a minimum at $y\neq 0$ for $r<1/3,$ so that the minimal value
of $\left\vert \overline{A}\right\vert $ is found at $q\neq 0$ for the total
thickness of the two dead layers exceeding the critical thickness $d_{m}$:
\begin{equation}
d>d_{m}=\epsilon _{e}\sqrt{\frac{3D}{\pi \epsilon _{\perp }}}  \label{dm}
\end{equation}%
This solution is:
\begin{eqnarray}
q_{c} &=&\frac{1}{l}\left( 15\frac{d-d_{m}}{d_{m}}\right) ^{1/2},
\label{eq:qc1} \\
k_{c} &=&\left( \frac{5\pi }{\epsilon _{e}}\frac{d-d_{m}}{Dl}\right) ^{1/2}.
\label{eq:kc1}
\end{eqnarray}%
{\LARGE \ }We see that for $d<d_{m}$ the function $\overline{A}\left(
q\right) $ is monotonous and has its maximum value at $q=0$, i.e. for very
thin dead layers the phase transition occurs into the homogeneously
polarized state. In the case $d\gtrsim d_{m},$ the phase transition occurs
into an inhomogeneously polarized state with $k=k_{c}$ given by Eq.(\ref%
{eq:kc1}), i.e. the \textquotedblleft wave vector\textquotedblright\ of the
sinusoidal polarization \textquotedblleft wave\textquotedblright\ increases
very rapidly from $k=0$ at $d=d_{m}$ to a large value and $ql/2$\ rapidly
becomes on the order of unity. The corresponding transition temperature is
found from
\begin{equation}
|\overline{A}|=\frac{4\pi d}{\epsilon _{e}l}\left( 1-\frac{5\left(
d-d_{m}\right) ^{2}}{4d_{m}^{2}}\right) ,
\end{equation}%
i.e. it occurs just slightly above the transition into the homogeneous
state, where $|\overline{A}_{h}|=4\pi d/\epsilon _{e}l.$ For BaTiO$_{3}$%
/SrRuO$_{3}$ the estimate gives a small value $d_{m}=0.3\mathring{A},$ hence
for BTO/SRO\ such a regime is only of an academic interest, $d\simeq 5d_{m}$
there (see Sec.VI). However, it seems to be very interesting to look for
systems where $d_{m}$ may be large, since those systems could perhaps
sustain the homogeneous polarization and, therefore, are most interesting
for memory applications.

In the opposite limiting case{\normalsize , }$d\gg d_{m}${\normalsize , }%
which is a typical real situation, $ql$ increases and approaches $\pi ,$ so
that the value of $\tan \frac{ql}{2}$ becomes very large. We suppose that in
this case $\epsilon _{\perp }k^{2}\gg \epsilon _{b}q^{2}$ and shall check
later that this assumption{\normalsize \ }is justified. Then,{\normalsize \ }%
one can omit $\epsilon _{b}q^{2}$ in the denominator of the first term on
r.h.s. of Eq.~(\ref{eq:Akq}). Also, the last term can be omitted, because,
as we shall see below, the maximum of $\overline{A}$ corresponds to $k\gg q$
and the coefficients $D$ and $\eta $ are normally of the same order of
magnitude. Indeed, the maximum of the simplified expression corresponds to
\begin{equation}
k=k_{c}=\left( \frac{4\pi q_{c}^{2}}{\epsilon _{\perp }D}\right)
^{1/4}=\left( \frac{4\pi ^{3}}{\epsilon _{\perp }Dl^{2}}\right) ^{1/4},
\label{eq:kc2}
\end{equation}%
since, as we argued above,
\begin{equation}
q_{c}\simeq \pi /l.  \label{eq:qc2}
\end{equation}%
In this regime $\epsilon _{\perp }^{1/2}${\normalsize \ }$k_{c}/q_{c}=\left(
4\epsilon _{\perp }/\pi \right) ^{1/4}l^{1/2}/D^{1/4}$. Given that usually $%
D $\ is on the order of magnitude of $d_{at}^{2},$\ where $d_{at}$\ is the%
{\normalsize \ }typical interatomic distance (cf. Appendix E) and%
{\normalsize \ }$\left( 4\epsilon _{\perp }/\pi \right) ^{1/4}${\normalsize %
\ }is on order of unity, in all real cases one has approximately{\normalsize %
: }$\epsilon _{\perp }^{1/2}k_{c}/q_{c}\simeq \left( l/d_{at}\right) ^{1/2},$%
\ i.e. our assumption that{\normalsize \ }$\epsilon _{\perp }k^{2}\gg
\epsilon _{b}q^{2}${\normalsize \ }for{\normalsize \ }$d\gg d_{m}$%
{\normalsize \ }is justified for films thicker than several monolayers.

We see that with increasing thickness of the dead layer $d$ the period of
the sinusoidal domain structure rapidly decreases and then saturates at a
value \emph{independent} of $d$. Since at $d-d_{m}=d_{m}$ both Eq.~(\ref%
{eq:kc1}) and Eq.~(\ref{eq:kc2}) give similar values for $k_{c}$ (difference
is about $20\%$), the two simple formulas are sufficient to describe the
function $k_{c}\left( d\right) $ with a reasonable accuracy at all values of
$d$. From Eqs.~(\ref{eq:kc2}), (\ref{eq:qc2}), and (\ref{eq:Akq}), one sees
that at $d-d_{m}>d_{m}$ the loss of stability of the paraelectric phase with
respect to formation of the domain structure takes place at
\begin{equation}
\overline{A}=-2Dk_{c}^{2}.  \label{loss}
\end{equation}%
The relation (\ref{eq:Akq})\ allows to simplify the expressions for the
inhomogeneous fields in the ferroelectric film (\ref{Ezf}),(\ref{Exf}),
which we rewrite, using $\epsilon _{\perp }k^{2}\gg \epsilon _{b}q^{2}$, as:
\begin{eqnarray}
\widetilde{E}_{fz}\left( x,z\right) &=&-a\frac{4\pi ^{3}}{\epsilon _{\perp
}\left( kl\right) ^{2}}\cos kx\cos qz,  \label{eq:Efz2} \\
\widetilde{E}_{fx}\left( x,z\right) &=&a\frac{4\pi ^{2}}{\epsilon _{\perp }kl%
}\sin kx\sin qz.  \label{eq:Efx2}
\end{eqnarray}%
Obviously, both fields are much smaller than the magnitude of the
polarization wave:\ since $kl\sim \sqrt{l/d_{at}}\gg 1$, then $\left\vert
\widetilde{E}_{fz}\right\vert \ll \left\vert \widetilde{E}_{fx}\right\vert
\ll a.$\

\begin{figure}[h]
\begin{centering}
\includegraphics [width=6.5cm]{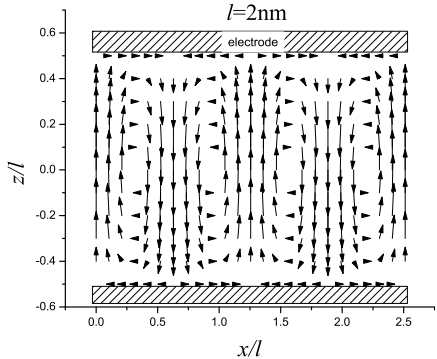}
\caption{ The polarization distribution in the BTO/STO film with
thickness $l=2$nm in sinusoidal domain regime. It clearly shows that
the electric field component along the film near its surface results
in the presence of perpendicular to the easy $z-$axis component of
the polarization in the narrow sinusoidal domains. The pattern is
similar to the one for the  structure with abrupt narrow domains
(the Kittel structure). } \label{fig:Pvortmap}
\end{centering}
\vspace{0.05 cm}
\end{figure}

In comparison, the field in the dead layers in the studied case $kd\ll 1$
is, near FE interface:
\begin{eqnarray}
\widetilde{E}_{dz}\left( x,z=\frac{l}{2}\right) &=&a\frac{8\pi ^{2}}{%
\epsilon _{\perp }k^{2}ld}\cos kx  \notag \\
&=&a\frac{4\pi ^{1/2}D^{1/2}}{\epsilon _{\perp }^{1/2}d}\cos kx,
\label{eq:Etdz} \\
\widetilde{E}_{dx}\left( x,z=\frac{l}{2}\right) &=&a\frac{4\pi ^{2}}{%
\epsilon _{\perp }kl}\sin kx,  \label{eq:Etdx}
\end{eqnarray}%
hence, $\widetilde{E}_{dz}$ is much larger than any other field component in
the system. As we can see from (\ref{eq:Etdz}), there is no small parameter
like $d_{at}/l$ or $d/l$ in the $z-$component of the field in the dead layer
$\widetilde{E}_{dz}$ (the only parameter there is $d_{at}/d,$ $d_{at}\sim
D^{1/2}$ [see discussion after Eq.(\ref{eq:qc2})], which is less than unity,
yet much larger than the above mentioned other small parameters in the
problem.) This is reminiscent of the behavior that we saw above for the case
of the homogeneous polarization. These results allow us to construct the
maps of the polarization $\widetilde{P}_{z}$ (\ref{Ansatz1}) and $\widetilde{%
P}_{x}=(\epsilon _{\perp }-1)\widetilde{E}_{fx}/4\pi $ characteristic of the
sinusoidal polarization distribution, shown in Fig.~\ref{fig:Pvortmap}. The
map shows the narrow sinusoidal domains with well ordered polarization
distribution. The stray electric field component along the film near its
surface, $\tilde{E}_{x}$, results in the presence of perpendicular to the
easy $z-$axis component of the polarization, $P_{x}$, in the narrow
sinusoidal domains. The pattern is similar to the one for the structure with
abrupt domains (the well-known Kittel structure with thin domain walls\cite%
{kittel}.)

\subsection{Metallic electrode}

We would like to compare now the ferroelectric film with the dead layers to
the one of real metallic electrodes with Thomas-Fermi screening, in order to
see the analogy between the two cases. Obviously, the inhomogeneous
polarization in the form (\ref{Ansatz1}) will produce the electric field and
the electrostatic potential $\varpropto \cos kx,$ so that the electrostatic
potential will be $\phi (z,x)=\Phi (z)\cos kx.$ The Poisson equation then
takes the form
\begin{equation}
\left( \partial _{z}^{2}-k^{2}\right) \Phi =\Phi /\lambda ^{2},
\end{equation}%
that has the solution $\Phi (z)=-\Phi (-z):$
\begin{eqnarray}
\Phi &=&\phi _{1}e^{-\kappa (z-l/2)},\qquad z>l/2, \\
\kappa &=&\sqrt{k^{2}+\lambda ^{-2}}.
\end{eqnarray}%
With this solution, the boundary conditions, found analogously to (\ref{bc1}%
),(\ref{bc2}), give:
\begin{eqnarray}
-a\frac{k}{q}\left( \overline{A}+Dk^{2}+\eta q^{2}\right) \sin \frac{ql}{2}
&=&k\phi _{1},  \label{eq:bcm1} \\
a\left( \overline{A}+4\pi /\epsilon _{b}+Dk^{2}+\eta q^{2}\right) \cos \frac{%
ql}{2} &=&\frac{\epsilon _{e}}{\epsilon _{b}}\kappa \phi _{1}.
\label{eq:bcm2}
\end{eqnarray}%
Excluding $\overline{A}$ from these equations with the help of Eq.(\ref%
{eq:Akq}), one arrives at \cite{BLapl06}:
\begin{equation}
\frac{q}{\epsilon _{\perp }}\tan \frac{ql}{2}=\frac{k^{2}}{\epsilon _{e}%
\sqrt{k^{2}+\lambda ^{-2}}}.  \label{eq:kqmet}
\end{equation}%
Considering the case of interest to us, $k\lambda \ll 1,$ and comparing this
with (\ref{kq2}), we see that for $\epsilon _{\perp }k^{2}\gg q^{2}$ the
case of realistic metallic electrodes and the simple dead layer model are
mathematically equivalent when $\lambda =d/2.$ Since $d/2$ is the thickness
of the dead layer, the result suggests the numerical equivalence of the dead
layer and the Thomas-Fermi screening length when $\epsilon _{\perp }k^{2}\gg
q^{2}$.

\section{Properties of a Sinusoidal Domain Structure}

Now we turn to the discussion of the ferroelectric phase. It is quite
natural to expect that close to the transition the spatial distribution of
the polarization may be well represented by a single or a few sinusoids with
the wave vector equal to $k_{c}$ in addition to a spatially homogeneous part
$p$. There will certainly be a single sinusoidal wave, at least close to the
transition, if there is an anisotropy in the ($x-y$) plane. In this review,
we discuss this latter case only.

\subsection{Range of existence}

Consider first a zero external field, $E_{0}=0$, when the homogeneous part
of the polarization is zero both in the paraelectric and the ferroelectric
phases, i.e. close to the transition there is a single sinusoid and $p=0$.
Some distance away from the transition, the single sinusoidal approximation (%
\ref{Ansatz1}) breaks down. Indeed, because of the cubic term in Eq.~(\ref%
{eqstate1a}), the polarization of the form $\widetilde{P}=a\cos kx\cos qz$
suggests a presence of an electric field, which depends on the coordinates
not only as $\cos kx\cos qz,$ but also as $\cos 3kx\cos qz,$ $\cos kx\cos
3qz,$ $\cos 3kx\cos 3qz.$ Such a field induces polarization containing even
higher spatial harmonics, and so forth ad infinitum. Let us estimate when
these higher contributions to the field and the higher harmonics in the
polarization may be neglected. Suppose that the main part of the
polarization is the simplest polarization wave (\ref{Ansatz1}). When the
polarization is known, one can calculate the electric field produced by this
polarization without referring to the equation of state. In our case, it is
convenient to use the relation between the polarization and the field at the
transition point given by Eq.~(\ref{eq:Etz}), which yields:
\begin{eqnarray}
\widetilde{E}_{fz} &=&-\frac{4\pi q_{c}^{2}}{\epsilon _{\perp }k_{c}^{2}}%
a\cos k_{c}x\cos q_{c}z  \notag \\
&=&-Dk_{c}^{2}a\cos k_{c}x\cos q_{c}z,
\end{eqnarray}%
since $\epsilon _{\perp }k^{2}\gg q^{2}$. It is impossible to satisfy
exactly the equation of state (\ref{eqstate1a}) with such a field and
polarization, and we have to consider the corrections to the field $\check{E}%
_{z}\left( x,z\right) $: $E_{z}=\widetilde{E}_{fz}+\check{E}_{z}.$ Let us
try $\widetilde{P}_{z}$ in the form (\ref{Ansatz1})\ in (\ref{eqstate1a})\
to see what is lacking. We obtain:
\begin{eqnarray}
&&-Dk_{c}^{2}a\cos k_{c}x\cos q_{c}z+\check{E}_{z}\left( x,z\right)  \notag
\\
&=&\left( A+Dk_{c}^{2}+\frac{9Ba^{2}}{16}\right) a\cos k_{c}x\cos q_{c}z
\notag \\
&&+\frac{Ba^{3}}{16}(3\cos 3k_{c}x\cos q_{c}z  \notag \\
&&+\cos k_{c}x\cos 3q_{c}z+\cos k_{c}x\cos 3q_{c}z).
\end{eqnarray}%
One can satisfy this equation by putting
\begin{equation}
A+2Dk_{c}^{2}+\frac{9}{16}Ba^{2}=0,
\end{equation}%
or
\begin{equation}
a^{2}=-\frac{16}{9}\frac{A+2Dk_{c}^{2}}{B}=-\frac{16}{9}\frac{\widetilde{A}}{%
B},
\end{equation}%
and introducing the additional \textquotedblleft ghost\textquotedblright\
fields in order to compensate the rest of the terms on the right hand side.
They are not the real electric fields because they do not satisfy the
electrostatic equations. In fact, it is inconsistent to take into account
the higher harmonics of the field without correcting the polarization
self-consistently. But all this trouble is not necessary if those
\textquotedblleft ghost fields\textquotedblright\ are small compared to the
real ones. The condition for this reads
\begin{equation*}
Dk_{c}^{2}>\frac{3}{16}Ba^{2}=-\frac{1}{3}\widetilde{A},
\end{equation*}%
or
\begin{equation}
-\widetilde{A}<3Dk_{c}^{2}.  \label{condappl}
\end{equation}%
The numerical factor is certainly not reliable and may be omitted.

If the polarization is presented not as a single sinusoid but as a sum of a
constant and a sinusoidal parts the condition of neglecting of the higher
order harmonics has to be considered anew. In this case, we have close to
the phase transition:
\begin{equation}
P_{z}(x,z)=p+a\cos k_{c}x\cos q_{c}z.  \label{Ansatz2}
\end{equation}%
For $a,$ we now have from the equation of state,
\begin{equation}
a^{2}=-\frac{16}{9}\frac{A+2Dk_{c}^{2}+3Bp^{2}}{B},  \label{aeq}
\end{equation}%
and among the \textquotedblleft ghost" fields there is, for example, the
component $\frac{3}{4}Bpa^{2}\cos 2k_{c}x\cos 2q_{c}x$ that can be neglected
when
\begin{equation}
\frac{4\pi q_{c}^{2}}{\epsilon _{\perp }k_{c}^{2}}>\frac{3}{4}Bpa=Bp\sqrt{-%
\frac{\widetilde{A}+3Bp^{2}}{B}}.
\end{equation}%
The maximum of the r.h.s. is at $p=\sqrt{-\widetilde{A}/6B\text{ }},$ so
this condition transforms into:
\begin{equation}
-\widetilde{A}<2\sqrt{3}Dk_{c}^{2},
\end{equation}%
which is practically the same as (\ref{condappl}). Let us emphasize that
\emph{at any temperature} the same formula may be applied to describe the
states with $p_{c}\lesssim p,$ since the amplitude $a$ cannot be large in
this case.

\subsection{Free energy}

It was convenient and instructive to use the equations of state to describe
instability of the paraelectric phase. However, it is more convenient to use
a (non-equilibrium) free energy in order to analyze the properties of the
sinusoidal domain structures, whose minimization provides us with an
equilibrium state. The LGD free energy (\ref{LGD}) does not include the
energy of the electric field sources, and we shall calculate it now. But
first, a simple exercise: let us perform a similar calculation for a linear
dielectric with the ideal metallic electrodes at a non-zero voltage $U.$ For
simplicity, we assume that the dielectric is extremely anisotropic and has
one component of the polarization only, and that only the states with a
homogeneous polarization are studied. We have
\begin{equation}
\tilde{F}=\int dV\frac{A}{2}P^{2}+W_{el},  \label{eq:Flind}
\end{equation}%
where the integral is taken over the volume of the dielectric and $W_{el}$
is the electric energy. The integrand can be called the LGD free energy for
the linear dielectric. The electric energy can be represented as
\begin{equation}
W_{el}=\int dV\frac{E^{2}}{8\pi },
\end{equation}%
where the integral is now over \textquotedblleft the whole
universe\textquotedblright , which means in our case over the volume of the
dielectric and the voltage source. The latter is imagined as a capacitor
with an infinite capacitance. The electric field in the dielectric is always
$U/l$ and does not depend on the polarization. When we change the
polarization (recall that we consider the \emph{non-equilibrium} free
energy), the charge on the electrodes is changed and, therefore, the charge
and the energy of the source are changing too. Suppose first that the
polarization is zero and for this state of the dielectric the charge of the
source is $Q_{s},$ so its energy is equal to $Q_{s}^{2}/2C$, where $C$ is
the capacitance of the source. Consider now a state with the polarization $%
P. $ The charge on the electrode with the potential $U$ (the second
electrode has zero potential by assumption) has then changed by some value $%
Q $. The electric energy of the source has changed by $\left[ \left(
Q_{s}-Q\right) ^{2}-Q_{s}^{2}\right] /2C=-QU$, where we have taken into
account that $U=Q_{s}/C$ and that $Q^{2}/2C\rightarrow 0$ when $C\rightarrow
\infty $. In general, the work of the voltage source (change of its energy)\
can be written as
\begin{eqnarray}
-\sum_{a=1,2}^{\mathrm{electrodes}}e_{a}\varphi _{a} &=&-(e_{1}\varphi
_{1}+e_{2}\varphi _{2})=-e_{1}\varphi _{1}  \notag \\
&=&-QU,
\end{eqnarray}%
where $e_{a}$ are the charges on the electrodes, and we assumed that the
second electrodes in earthed. When the polarization is directed outward from
the electrode, then the charge on the plate is $Q=PS,$ where $S$ is the
electrode (plate) area. Since the electric energy lowers in this case, the
equilibrium polarization will be directed from the electrode $U.$ We have
proven a well known fact that the polarization tends to be oriented along
the field. Now, the full change of the energy per unit area due to a
non-zero polarization is $\tilde{F}/S=lAP^{2}/2-PU,$ where $l$ is the
thickness of the dielectric plate. Minimizing this non-equilibrium free
energy with respect to $P,$ we find $P=\frac{1}{A}\frac{U}{l}.$ This is, of
course, a well-known result for a linear dielectric: the equilibrium
polarization is directed along the electric field and is proportional to the
the field (cf. the standard $P=\chi E)$. Using this, one can find that the
sum of the \textquotedblleft LGD\textquotedblright\ energy of a linear
dielectric (\ref{eq:Flind})\ and the energy of electric field inside it at
the equilibrium is:
\begin{equation}
\int \left( \frac{A}{2}P_{z}^{2}+\frac{E^{2}}{8\pi }\right) dV=\int \frac{%
\epsilon E^{2}}{8\pi }dV,  \label{elendiel}
\end{equation}%
where $\epsilon =1+4\pi \chi ,$ which is another well known formula.

Now we can evaluate the free energy of a ferroelectric capacitor at a given
bias voltage. Since we are interested in the case of an inhomogeneous
ferroelectric polarization, we have to take into account the polarization
perpendicular to the polar axis, cf. Eq.~(\ref{Exf}). For the LGD energy of
this polarization together with the energy of the perpendicular electric
field in the ferroelectric, we can use Eq.~(\ref{elendiel}). The same is
valid for the nonferroelectric part of polarization of the ferroelectric
along axis $z$, as well as for the polarization and the electric field of
the dead layer. As a result, we obtain a contribution to the full free
energy:
\begin{equation}
\int_{FE}\frac{\epsilon _{\perp }E_{\perp }^{2}}{8\pi }dV+\int_{FE}\frac{%
\epsilon _{b}E_{z}^{2}}{8\pi }dV+\int_{DL}dV\frac{\epsilon _{e}E^{2}}{8\pi },
\end{equation}%
where the first integral is over the volume of the ferroelectric (FE)\ and
the second over the volume of the dead layer (DL). It is convenient, though
not quite consistent, to refer to these contributions as to {\Large \ }the
electric field energy

The total energy consists, therefore, of three parts: the LGD\ contribution,
the electric field energy and the energy of the voltage source. It can be
presented as (here and below, we will always imply that the free energy is
calculated per unit area)
\begin{align}
\tilde{F}& =\int_{FE}dV\Bigl[\frac{A}{2}P_{z}^{2}+\frac{B}{4}%
P_{z}^{4}+\ldots +\frac{1}{2}D_{ij}\left( \nabla _{\bot i}P_{z}\right)
\left( \nabla _{\bot j}P_{z}\right)  \notag \\
& +\frac{1}{2}\eta \left( \partial _{z}P_{z}\right) ^{2}  \notag \\
& +\frac{\epsilon _{\perp }E_{\perp }^{2}}{8\pi }+\frac{\epsilon
_{b}E_{z}^{2}}{8\pi }\Bigr]+\int_{DL}dV\frac{\epsilon _{e}E^{2}}{8\pi }%
-\sum_{a}^{\mathrm{electrodes}}e_{a}\varphi _{a},
\end{align}%
where the first integral is over the volume of the ferroelectric (FE)\ and
the second over the volume of the dead layer (DL). At the moment, we want to
study an anisotropic film where the polarization changes only along one of
the transversal axes (e.g. $x$) and the standard case of $q\ll k$ (see our
discussion above). Then, we can simplify the free energy as:
\begin{eqnarray}
\tilde{F} &=&\int_{FE}dV\Bigl[\frac{A}{2}P_{z}^{2}+\frac{B}{4}P_{z}^{4}+%
\frac{1}{2}D\left( \partial _{x}P_{z}\right) ^{2}+\frac{\epsilon
_{b}E_{z}^{2}}{8\pi }+\frac{\epsilon _{\perp }E_{x}^{2}}{8\pi }\Bigr]  \notag
\\
&&+\int_{DL}dV\frac{\epsilon _{e}E^{2}}{8\pi }-QU.  \label{eq:Ftil}
\end{eqnarray}%
Let us mention that the electric field energies for the homogeneous and the
sinusoidal parts are additive, as follows from the fact that $\int dx\cos
kx=0$\ .

Consider the electric field energy for the sinusoidal part.{\normalsize \ }%
We find from Eqs.~(\ref{Ezf}) and (\ref{Exf}), that
\begin{equation}
\widetilde{E}_{fz}\ll \widetilde{E}_{fx}=\frac{4\pi ^{2}a}{\epsilon _{\perp
}k_{c}l}\sin k_{c}x\sin q_{c}z,
\end{equation}%
hence, since $\epsilon _{b}$ is hardly larger than $\epsilon _{\perp }$, the
energy of the electric field in the ferroelectric is
\begin{equation}
\int \frac{\epsilon _{\perp }}{8\pi }\widetilde{E}_{xf}^{2}dV=l\frac{\pi
q_{c}^{2}}{2\epsilon _{\perp }k_{c}^{2}}a^{2}.
\end{equation}%
We obtain the field in the dead layer from Eq.~(\ref{bc1}). Recalling that
we are interested in the case $kd\ll 1,$ $q\ll k$ and $q\simeq \pi /l,$ we
find that in the dead layer:
\begin{equation}
\widetilde{E}_{dx}\ll \widetilde{E}_{dz}=\frac{8\pi ^{2}}{\epsilon _{\perp
}k_{c}^{2}ld}a\cos k_{c}x=\frac{4\pi ^{1/2}D^{1/2}}{\epsilon _{\perp }^{1/2}d%
}a\cos k_{c}x.
\end{equation}%
The energy of the electric field in the dead layer is
\begin{equation}
\epsilon _{e}\frac{\widetilde{E}_{dz}^{2}}{8\pi }d\left\langle \cos
^{2}k_{c}x\right\rangle =\frac{\epsilon _{e}D}{4\epsilon _{\perp }d}a^{2}=%
\sqrt{\frac{\pi }{3}}\frac{d_{m}}{d}\frac{D^{1/2}}{4\epsilon _{\perp }^{1/2}}%
a^{2}.
\end{equation}%
Taking into account Eqs.~(\ref{eq:kc2}), (\ref{eq:qc2}), one sees that the
ratio of the energies of the electric field in the ferroelectric and in the
dead layer is $\pi \sqrt{3}d/d_{m},$ i.e. for our case $d\gg d_{m}$ the
contribution of the deal layer can be neglected.

Now, it remains to calculate the contribution of the homogeneous part%
{\normalsize \ }of the electric field energy . To this end, we use Eq.~(\ref%
{extdep}) for the electric field in the ferroelectric, find the electric
field in the dead layer from Eq.~(\ref{eq:E0}), and obtain:
\begin{equation}
W_{el}=\frac{\epsilon _{b}l}{8\pi }E_{fz}^{2}+\frac{\epsilon _{e}d}{8\pi }%
E_{dz}^{2}=\frac{2\pi p^{2}dl}{l\epsilon _{e}+\epsilon _{b}d}+\frac{\epsilon
_{b}\epsilon _{e}U^{2}}{8\pi \left( \epsilon _{e}l+\epsilon _{b}d\right) }
\end{equation}%
{\LARGE \ }The last term here is the unimportant constant. Finally, the
energy of the external voltage source is $-QU$ and for a unit area one finds
from Eq.~(\ref{extdep}):
\begin{equation}
Q=\frac{l}{l+\epsilon _{b}d/\epsilon _{e}}p,
\end{equation}%
and
\begin{equation}
W_{els}=-\frac{l}{l+\epsilon _{b}d/\epsilon _{e}}pU=-lpE_{0}.
\end{equation}%
The total electric energy per unit volume equals
\begin{equation}
\frac{\pi q_{c}^{2}}{2\epsilon _{\perp }k_{c}^{2}}a^{2}+\frac{2\pi p^{2}dl}{%
l\epsilon _{e}+\epsilon _{b}d}-pE_{0}.
\end{equation}%
Finally, one obtains the LGD contribution to the free energy by using Eq.~(%
\ref{Ansatz2}) and calculating the integrals for $q=\pi /l.$ As a result,
the free energy per unit area $\tilde{F}(p,a)$ takes the form:
\begin{equation}
l^{-1}\tilde{F}(p,a)=\frac{\tilde{A}+\xi }{2}p^{2}+\frac{\tilde{A}}{8}a^{2}+%
\frac{B}{4}p^{4}+\frac{3B}{8}a^{2}p^{2}+\frac{9B}{256}a^{4}-pE_{0},
\label{eq:Fpa}
\end{equation}%
where
\begin{eqnarray}
\widetilde{A} &=&A+2Dk_{c}^{2},  \label{eq:At} \\
\xi &=&4\pi d/\left( \epsilon _{e}l+\epsilon _{b}d\right)
-2Dk_{c}^{2}\approx 4\pi d/\left( \epsilon _{e}l\right) -2Dk_{c}^{2},
\label{eq:xi}
\end{eqnarray}%
are the main parameters of the system, $E_{0}=\epsilon _{e}U/\left( \epsilon
_{e}l+\epsilon _{b}d\right) \approx U/l$ the external field for the usual
case of a thin dead layer $\epsilon _{b}d\ll \epsilon _{e}l.$ The Landau
coefficients $A,B,...$ in (\ref{LGD})\ are renormalized by lattice misfit
\cite{pertsev98} (see Appendix D), but we do not account for inhomogeneous
strain coupled with inhomogeneous polarization in Eq.~(\ref{eq:Ftil}). Its
account is irrelevant when one defines the point of stability of the
paraelectric phase with respect to the domain formation. Indeed, the
coupling between the inhomogeneous polarization and strains is nonlinear,
while the problem is linear. This coupling begins to play a role when the
finite amplitude $a$ of the \textquotedblleft polarization
waves\textquotedblright\ has to be found. Nevertheless, we shall not take it
into account to avoid unnecessary complications, which do not change our
main conclusions \cite{BLcomPer07}.

\section{Electric properties of sinusoidal domain structures}

\subsection{Polarization - External field curves. Different critical
thicknesses for stability loss and for memory}

By now, we have made an important step of calculating the free energy of the
film with the dead layers in a single harmonic approximation. Already at
this level, we can study the regions of (meta)stability of the homogeneous
state with respect to domains and determine the corresponding conditions. As
we have noted a several times above, the residual depolarizing field in the
ferroelectric resulting from an incomplete screening by the electrodes tends
to split FE into domains, and that defines the properties of very thin
films. Importantly, not only formation but also a response of the domain
structure to an external bias voltage appears to be affected by this field
by a much greater extent than one might have expected\cite{BLapl06}, as we
will discuss below. For instance, the $p-E_{f}$ dependencies have an unusual
shape with a negative slope, predicted earlier by the continuum theory for a
similar case at low temperatures\cite{BLPRB01}.

Apparently, the thin ferroelectric films are very well suited for an
application of the analytical theory of domain structures in ideal crystals.
The theory helps to understand some experimental data, as we shall
demonstrate in the next Section. However, at the moment this theory is
unable to make practical predictions that seem most important, like the
estimates of the parameters of the ferroelectric films that may correspond
to an acceptable memory performance. This is not surprising, however, since
there is no consistent theory of the FE switching.

In this regard, it is worth mentioning that the usual practice of using the
Kolmogorov-Avrami approach to describe the switching in FE is very
unfortunate. Such an approach was designed for considering nucleation and
growth in the absence of any long range interaction between the nuclei,
which is far from being the case in the ferroelectrics, and without account
for the depolarizing field suppressing creation of isolated nuclei. In other
words, strong macroscopic Coulomb interaction defining the behavior of thin
films is completely neglected in that approach, and there is no easy way to
take it into account. On the other hand, considering domain structures as a
superpositions of several sinusoidal distributions of polarization takes
consistently into account the Coulomb interaction from the very beginning.
Such a superposition is a readily identifiable periodic domain structure for
an equilibrium or a metastable state. Considering the switching, one most
likely should consider a local perturbation in the periodic structure
(nuclei), but it is very inconvenient in terms of the delocalized sinusoidal
basis functions. Nevertheless, discussion of thin films with nearly
sinusoidal domains gives a possibility to approach the problem of switching
from a different perspective.\ We shall see that even the first step in this
direction, i.e. considering purely sinusoidal domain structure with the use
of Eq.(\ref{eq:Fpa}) leads to the non-trivial conclusions. We believe that
further description of domain structures in this field is promising and
highly desirable.

Here, we present an analysis of the ferroelectric film with the sinusoidal
domain structure in the single harmonic approximation and will carefully
indicate where this approximation works. By minimizing the corresponding
free {\LARGE \ }energy (\ref{eq:Fpa}) with respect to $a,$ one finds the
equilibrium value of the amplitude:
\begin{eqnarray}
a_{0} &=&\left[ -\frac{16}{9B}\left( \tilde{A}+3Bp^{2}\right) \right] ^{1/2}=%
\frac{4p_{c}}{3^{1/2}}\sqrt{1-s^{2}}\text{ for }\left\vert s\right\vert <1,
\label{P0eq} \\
a_{0} &=&0\text{ for }\left\vert s\right\vert \geq 1,
\end{eqnarray}%
where $s=p/p_{c},$ with $p_{c}=\sqrt{-\tilde{A}/3B}$ the characteristic
polarization. Substituting this solution into Eq.(\ref{eq:Fpa}) we then
arrive at the \emph{dimensionless} free energy: $f=3B\xi l^{-1}\tilde{F}$
depending on the homogeneous polarization only through the parameter $s$:

\begin{equation}
f_{\pm }(s)=\left\{
\begin{array}{cc}
\frac{1}{2}y\left( 1-y\right) s^{2}+\frac{1}{12}y^{2}s^{4}-\sqrt{y}se, &
\left\vert s\right\vert \geq 1, \\
-\frac{y^{2}}{3}+\frac{y}{2}\left( 1+\frac{y}{3}\right) s^{2}-\frac{1}{4}%
y^{2}s^{4}-\sqrt{y}se, & \left\vert s\right\vert <1,%
\end{array}%
\right.  \label{eq:fpm}
\end{equation}%
where $e=E_{0}/\zeta $ is the relative external field, $\zeta =\xi ^{3/2}/%
\sqrt{3B}$ is the characteristic electric field, and
\begin{equation}
y=-\widetilde{A}/\xi ,  \label{eq:y}
\end{equation}%
is the characteristic temperature (i.e. the relative distance of transition
temperature from the paraelectric phase that depends on the film thickness $%
l $). One can easily see that the free energy is continuous with first
derivative with respect to $s,$ while $d^{2}f_{-}/ds^{2}\neq
d^{2}f_{+}/ds^{2}$ at $s=\pm 1$.

The equation of state $s=s\left( e\right) $ obtained \ from the condition $%
df/ds=0$ reads:
\begin{equation}
\sqrt{y}\left( 1-y\right) s+\frac{1}{3}y^{3/2}s^{3}=e,\qquad \left\vert
s\right\vert \geq 1  \label{eq:xplus}
\end{equation}
\begin{equation}
\sqrt{y}\left( 1+\frac{y}{3}\right) s-y^{3/2}s^{3}=e,\qquad \left\vert
s\right\vert <1,  \label{eq:xminus}
\end{equation}
and the given state is (meta)stable only when $d^{2}f/ds^{2}>0.$ It is very
instructive to compare the behavior of the FE capacitor with real electrodes
(or, equivalently, with the dead layers) with that of the system \emph{%
without} the dead layers, which does show the standard S-shaped polarization
loop described in the dimensionless units by
\begin{equation}
-\sqrt{y}\left( 1+y\right) s+\frac{y^{3/2}}{3}s^{3}=e.  \label{nodl}
\end{equation}

\begin{figure}[h]
\begin{centering}
\includegraphics [width=6cm]{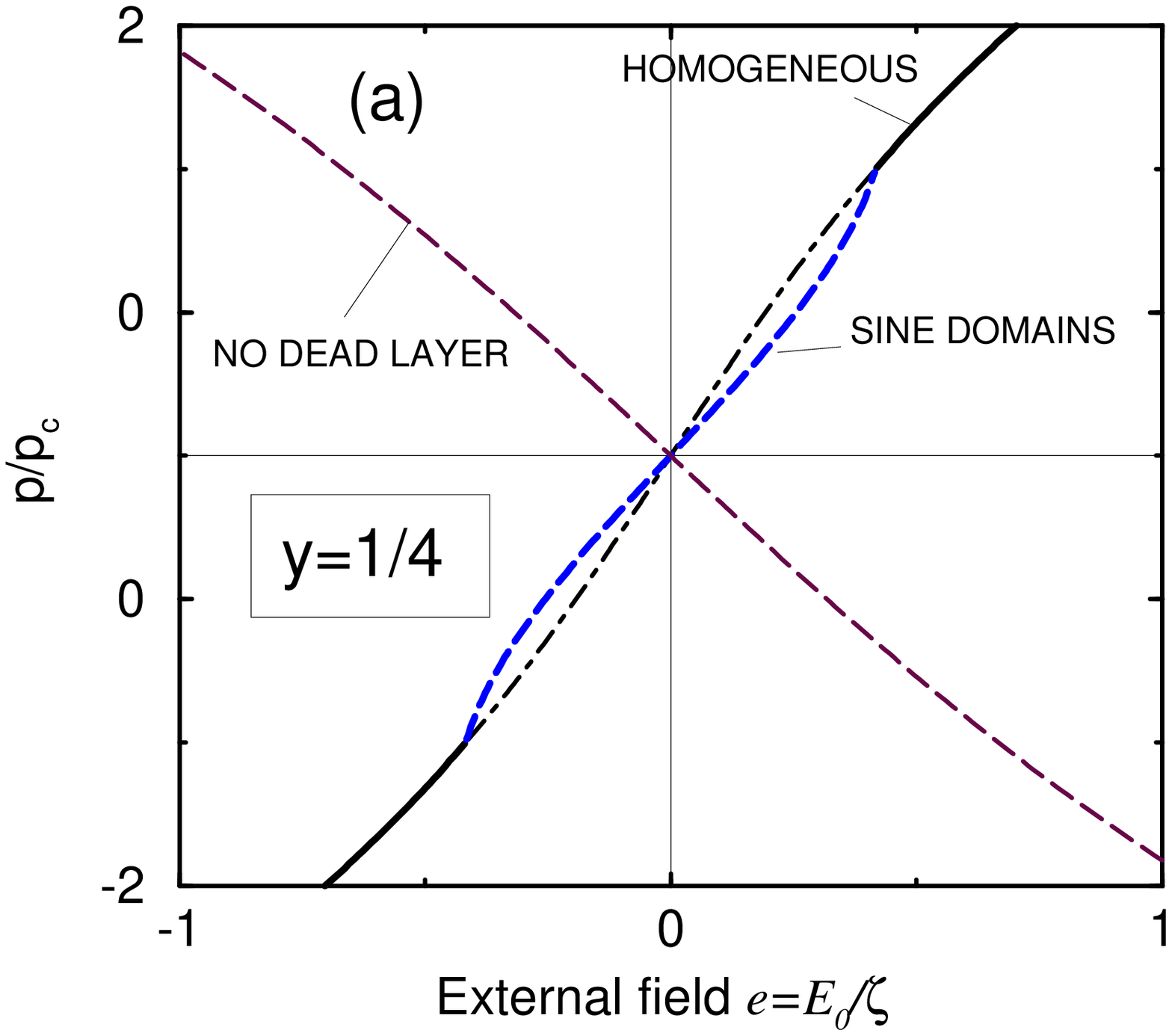}
\includegraphics [width=6cm]{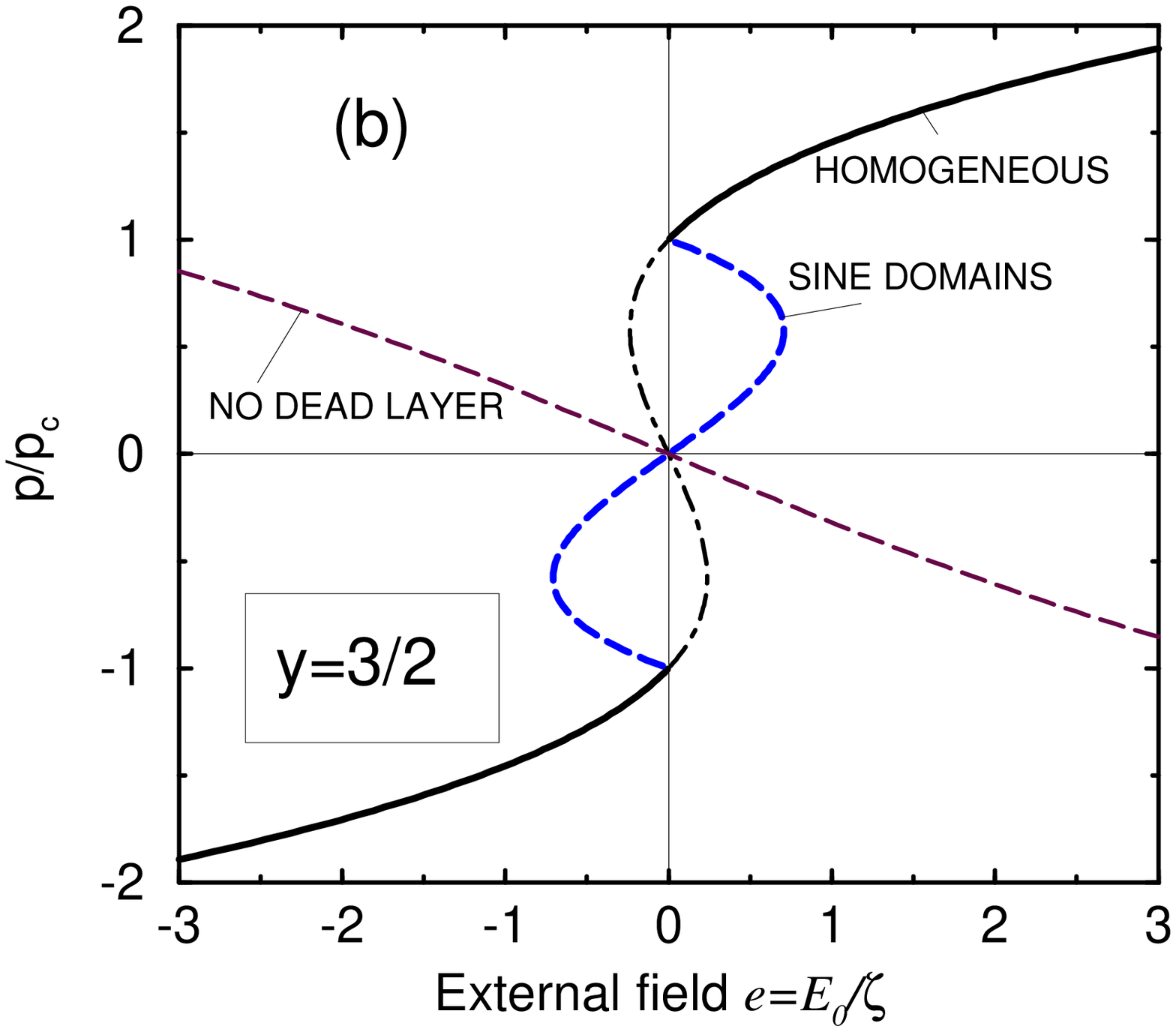}
\includegraphics [width=6cm]{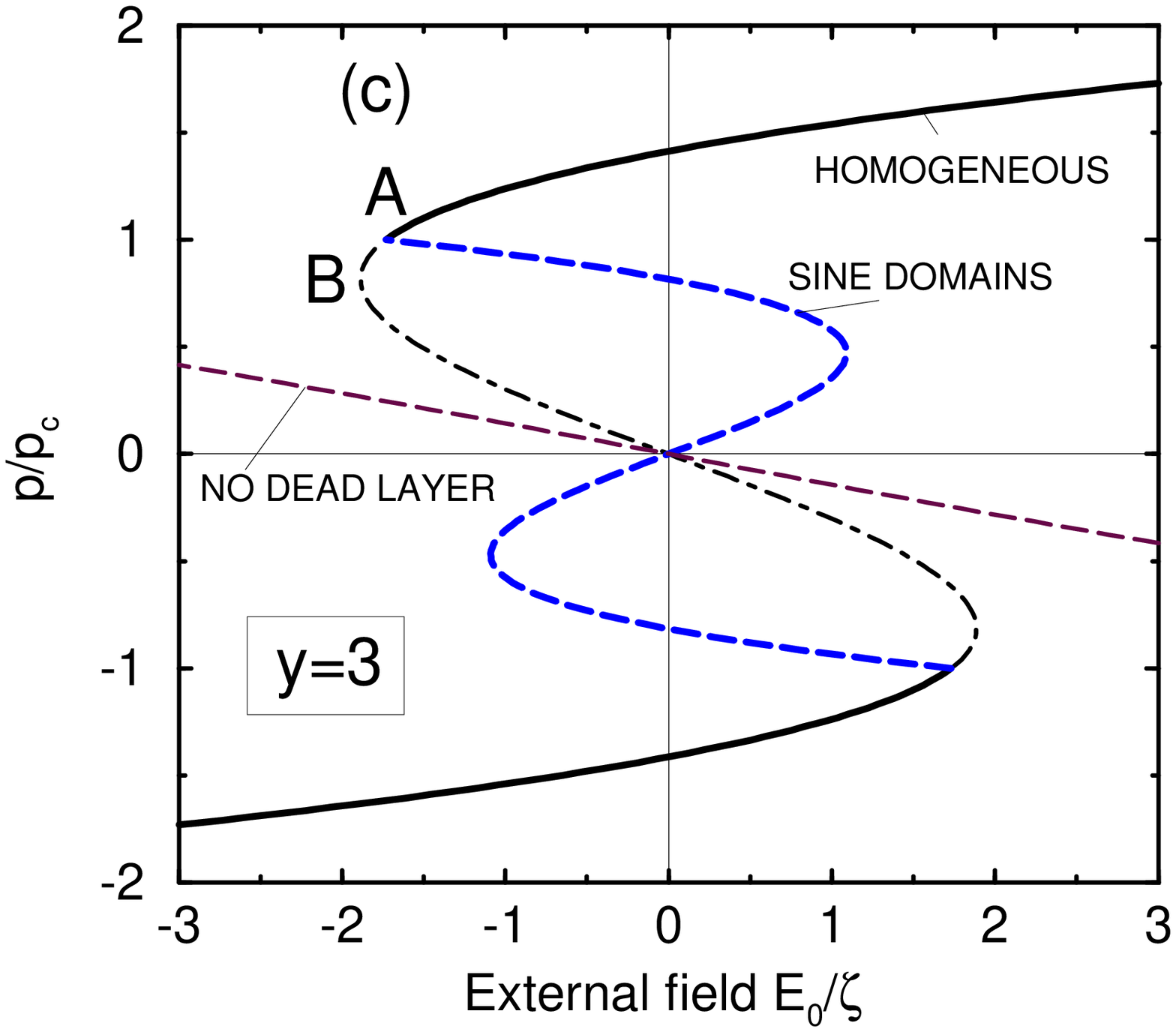}
\caption{The equation of state
$p=p(E_0)$ for ferroelectric film with dead layer in external field
$E_0$ for various values of the relative temperature $y$,
Eq.~(\ref{eq:y}): (a) $y=1/4$, (b) $y=3/2$, and (c) $y=3$. There is
a phase transition between homogeneous state and the one with
sinusoidal domains: (a) second order, which becomes first order in
cases (b) and (c). All those cases are very different from the case
of the FE film without the dead layer. $\zeta=\xi ^{3/2}/\sqrt{3B}$
is the characteristic electric field.}\label{fig:eqs}
\end{centering}
\vspace{0.05 cm}
\end{figure}

We illustrate the behavior of the equation of state depending on the
parameter $y$ of the system in Fig.~\ref{fig:eqs}. We have selected \textrm{%
\ }different values of $y=1/4,$\ $3/2,$\ and $3$, corresponding to different
temperatures (thus, $y=0$\ corresponds to the paraelectric-sinusoidal domain
structure transition for $U=0$\ occurring at $\tilde{A}=0$, at temperature $%
T_{d}<T_{c}$), with the results shown in Fig.~\ref{fig:eqs}.
We see that at a relatively small $y=1/4$ (for instance, for the system not
far below $T_{d}$)\ the equation of state (\ref{eq:xplus}), (\ref{eq:xminus}%
) has only a trivial solution $p=0$ at $E_{0}=0$, i.e. this is the state
with \emph{no memory}. In the field $e>e_{c}=\sqrt{y}-2y^{3/2}/3=5/12$ the
system is homogeneously polarized, it splits in the lower bias field via the
second order phase transition into domains with zero net polarization in the
film at $E_{0}=0.$ At $y=3/2$ (Fig.~\ref{fig:eqs}b) this transition is first
order (the change of the order of the phase transition occurs at $y=3/8$).
This is the first instance when the state with a spontaneous net
polarization $p\neq 0$ at $E_{0}=0$ becomes \emph{formally} possible as a
solution to the equations of state. However, this state is unstable ($%
d^{2}f_{-}/ds^{2}<0$ at $s=\pm 1),$ as is evident from a negative slope of $%
p-E_{0}$ $\ $curve for the upper branch of $f_{-}$ in Fig.~\ref{fig:eqs}b.

\begin{figure}[h]
\begin{centering}
\includegraphics [width=6cm]{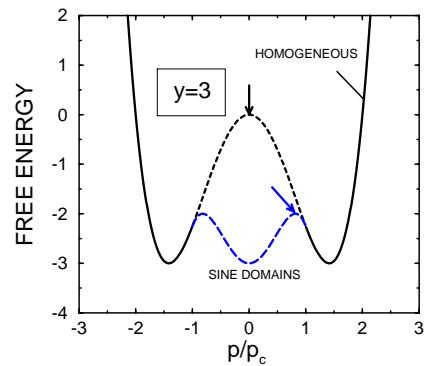}
\caption{The free energy of the FE
film with the dead layer for the relative temperature $y=3$, when
the transition between the homogeneously polarized state and
sinusoidal domains is first order. It is evident that switching
proceeds through the state with domains with much lower energy
barrier for nucleation (top of the barriers indicated by arrows). An
account for higher harmonics in Eq.~(\ref{Ansatz2}) will deepen the
energy of the domain state compared with the one shown here and will
lower the barrier even further.}\label{fig:Free}
\end{centering}
\vspace{0.05 cm}
\end{figure}

There is a \emph{metastability} of a \emph{homogeneously} polarized \emph{%
state} (we shall abbreviate it as the \textrm{MHS}) in the region $3/2<y<3$ [%
$d^{2}f_{+}/ds^{2}>0$ with the free energy higher than that of the domain
state, $f_{+}(\pm 1)>f_{-}(0)$ at $E_{0}=0$]. Formally, both the
ferroelectric memory and the polarization switching are possible at these
temperatures but no conclusion of practical importance can be made before
calculating the \emph{escape time} from the metastable state. At larger $%
y\geq 3$ (further down from the phase transition)\ the state with the
homogeneous polarization in the present single-harmonic approximation, Eq.~(%
\ref{Ansatz2}), has the same or lower free energy than the state with $p=0,$
$f_{+}(s_{m})\leq f_{-}(0),$ where $\pm s_{m}$ are the positions of the
minima of the free energy $f_{+}$ at $E_{0}=0$ (Fig.~\ref{fig:Free}).
However, this result is approximate. The reason is that Eq.~(\ref{Ansatz1})
is valid near the phase transition point only. The region of validity of
this approximation has been estimated in \cite{chensky82} and above (Sec.
IIIA){\normalsize \ }as roughly $-\tilde{A}<Dk_{c}^{2}$, which means $%
y\lesssim 1,$ at least for the experimental system BTO on SRO/STO considered
below in Sec. V.

In a more accurate approximation accounting for higher harmonics to describe
the inhomogeneous polarization, the free energy minimum at $s=0$ dips lower
than that of the homogeneous state. It is these higher harmonics that
convert the sinusoidal domain structure into a conventional one with narrow
domain walls. For the parts of the curves corresponding to $\left\vert
s\right\vert \gtrsim 1,$ these higher harmonics are not important (they are
when amplitude of the first harmonic becomes substantial) but they will
change the curves for $\left\vert s\right\vert <1$ substantially. The
amplitudes of the higher harmonics are to be considered as new variational
parameters for the free energy, and their account will be lowering the
estimated free energy. Hence, the minimum at $p=0$ in Fig.~\ref{fig:Free} is
actually deeper, and the homogeneously polarized phase becomes stable not at
$y=3$ but at a larger value (i.e. at a lower temperature or a \emph{larger
thickness}). Furthermore, it is possible that the homogeneous state would
always remain less stable than the polydomain state in that region. Indeed,
in the opposite limiting case, i.e. far below the FE transition, it has been
shown that for any thickness of the dead layer the multidomain state has a
lower free energy than the homogeneously polarized state\cite{BL2000}.

Importantly, the energy barrier between the oppositely polarized states is
\emph{strongly reduced} by the existence of the domain structure at $%
E_{0}\approx 0,$ Fig.~\ref{fig:Free}. It is very suggestive that one needs
to take this into account while considering the problem of polarization
switching or the problem of practical memory, but it is a problem that
should be dealt separately. We note again that Fig.~\ref{fig:eqs}c
highlights that homogeneous (single domain) switching is \emph{impossible}:
the homogeneously polarized phase loses its stability with respect to the
sinusoidal domain structure at the \emph{smaller} reverse bias (point A)
than the reverse bias flipping the homogeneous polarization (point B). The
two (necessarily negative)\ fields can be calculated from Eqs.~(\ref%
{eq:xplus}), (\ref{eq:xminus}). One finds the value of $e_{A}$\ either Eq.(%
\ref{eq:xplus}) or Eq.~(\ref{eq:xminus}) for $s=1$:
\begin{equation}
e_{A}=-\frac{2}{3}y^{3/2}+y^{1/2}  \label{eA}
\end{equation}%
The field $e_{B}$ can be obtained from Eq.~(\ref{eq:xplus}) by extrapolating
it to a region $s<1$ and calculating $s$\ and $e$\ corresponding to the
point $de/ds=0.$ One finds,
\begin{equation}
e_{B}=-\frac{2}{3}\left( y-1\right) ^{3/2}  \label{eB}
\end{equation}%
We see that the difference between the fields vanishes for $y\gg 1$, but it
is appreciable for small values of $y:$\ thus, for $y=2$\ the difference is
about 30\%. The applicability of Eqs.~(\ref{eA}),(\ref{eB}) is not limited
by the region of the one-sinusoidal approximation.

One may ask a question about what happens in a more realistic case where
higher powers of $P$ in the Landau expansion, $P^{6}$ or $P^{8},$ may be
important, like in BTO on STO\cite{BLapl06}. In this case the transition
seems to be second order according to first-principles calculations\cite%
{ghosez03}, same as in an earlier parameterization \cite{pertsev98}, while
the recent parameterizations \cite{LiCross05,tagBTO07} indicate very weak
first order phase transition. A full analysis is fairly involved in the case
of more general LGD functional and is beyond the scope of the present work.
Nevertheless, the condition for domain instability in a simplest
approximation takes the form
\begin{equation}
d^{2}F/dp^{2}=-2Dk_{c}^{2},  \label{eq:GenInstab}
\end{equation}
where $F$ is the LGD functional for the bulk (or a sample with ideal
metallic electrodes) and the second derivative is calculated for $p=p_{e},$
the homogeneous polarization in a given external field $E_{0}$ (this
condition will be modified by the strain, but it is reasonable to leave this
effect out as a first step). Our analysis indicates that the \emph{meta}%
stability of the homogeneous state now starts at a smaller $y,$ i.e. closer
to the point of the stability loss compared with $y>3/2$ value for the
second order case (\ref{eq:xminus}). However, since a first order transition
occurs before the stability loss, it is not clear if the temperature
interval between the transition into ferroelectric phase and appearance of
the MHS increases or decreases.

\subsection{Negative slope of $P(E_{f})$\ curves}

One unusual property accompanying the domain structure is the negative slope
of the $P=P(E_{f})$\ curves\ for ferroelectric\ films with domain structure.
Has been predicted some time ago for ideal films at low temperatures in Ref.
\cite{BLprlNegEf01,BLPRB01} and is viewed as a hallmark of a domain
structure governed mainly by electrostatics\cite{BLapl06}. Apparently, this
was first experimentally confirmed by the data from Noh \textit{et al.} \cite%
{KimL05}, who measured the hysteresis loops $P=P(E_{0})$ of ultrathin BTO
films on the SRO/STO\ substrate. It has been very instructive\cite{BLapl06}
to replot them as a function of a field in the ferroelectric $E_{f}$, $%
P=P(E_{f}).$ A remarkable specific feature of replotted loops is that they
all have a \emph{negative slope}, most pronounced at smaller film
thicknesses. Certainly, the theory does not give hysteresis loops since its
application is limited to an equilibrium state, but the behavior of the
equilibrium curve is clearly seen from the data\cite{BLapl06}.

We shall show below that the negative slope of the $P=P(E_{f})$\ is
characteristic of sinusoidal domain structure near the phase transition. It
is worth noting, however, that the negative slope $P/E_{f}<0$ is not an
exclusive property of the domain structure. It already happens in the
paraelectric phase in the temperature interval $-2Dk_{c}^{2}<A<0,$ where the
paraphase is still stable with regards to small inhomogeneous perturbations.
To see this for the sinusoidal domain structure, we express the external
field through the field in the ferroelectric $E_{f}$ with the use of (\ref%
{extdep}) the additional dimensionless parameter $\psi =1+\xi \left(
\epsilon _{e}l+\epsilon _{b}d\right) /4\pi d=\pi d_{m}/\sqrt{3}d$. We obtain:

\begin{equation}
E_{f}=E_{0}-\frac{4\pi dp}{\epsilon _{e}l+\epsilon _{b}d}=E_{0}-\frac{\xi p}{%
1-\psi }.  \label{eq:EfE0}
\end{equation}%
To find the response of the sinusoidal domains to the field in the
ferroelectric, $dp/dE_{f}$, it is most illustrative to use the equation of
state analogous to (\ref{eq:xminus})\ written in natural units, like
\begin{equation}
\left( -\frac{\widetilde{A}}{3}+\xi \right) p-3Bp^{3}=E_{0},\qquad
\left\vert p\right\vert <p_{c}.  \label{eq;stp}
\end{equation}%
Now, it is convenient to divide this equation by $Bp_{c}^{3}=-\widetilde{A}%
p_{c}$ and express $E_{0}$ through $E_{f}$ from Eq.~(\ref{eq:EfE0}). We can
then rewrite (\ref{eq:xminus}) as\
\begin{equation}
\left( \frac{1}{3}-\frac{\psi }{y(1-\psi )}\right) \frac{p}{p_{c}}-\left(
\frac{p}{p_{c}}\right) ^{3}=\frac{E_{f}}{3Bp_{c}^{3}}.
\end{equation}%
This yields for the slope at $E_{f}=0:$%
\begin{eqnarray}
\frac{dp}{dE_{f}} &=&-\frac{1}{\widetilde{A}}\frac{3y(1-\psi )}{y\left(
1-\psi \right) -3\psi }=\frac{1-\psi }{\xi }\frac{3}{y(1-\psi )-3\psi }
\notag \\
&=&\frac{3\epsilon _{e}l}{4\pi d}\frac{1}{y(1-\psi )-3\psi },
\end{eqnarray}%
where we have used $y=-\widetilde{A}/\xi ,$ and $\xi /(1-\psi )=4\pi
d/(\epsilon _{e}l+\epsilon _{b}d).$ Finally, this gives for the
\textquotedblleft dielectric constant" of the ferroelectric film :
\begin{equation}
\epsilon _{f}=1+4\pi dP/dE_{f}=\frac{3\epsilon _{e}l}{d}\frac{1}{y(1-\psi
)-3\psi }.  \label{eq:epsf}
\end{equation}%
At the point of stability loss with respect to sinusoidal structure, the
dielectric constant is already negative, as we mentioned above:
\begin{equation}
\epsilon _{f}=-\frac{3\epsilon _{e}l}{d\psi }=-\frac{4\pi }{2Dk_{c}^{2}}=-l%
\sqrt{\frac{\epsilon _{\perp }}{\pi D},}
\end{equation}%
This result is, of course, identical to the standard value of the
\textquotedblleft dielectric constant" of the FE\ film itself
\begin{equation}
\epsilon _{f}^{\mathrm{para}}=1+\frac{4\pi P}{E_{f}}\approx \frac{4\pi }{A}=-%
\frac{4\pi }{2Dk_{c}^{2}}<0,  \label{eq:enegP}
\end{equation}%
at the border of stability loss into sinusoidal domain structure. Note that
the negative sign of $\epsilon _{f}$ does not mean an instability, because
this coefficient does not characterize the response of the system as a whole
to an external field, which is $E_{0},$\ not $E_{f}.$\ The capacitance of
the device is defined as a response to $E_{0}$\ that is always positive\cite%
{BLPRB01}.

\begin{figure}[tbp]
\includegraphics{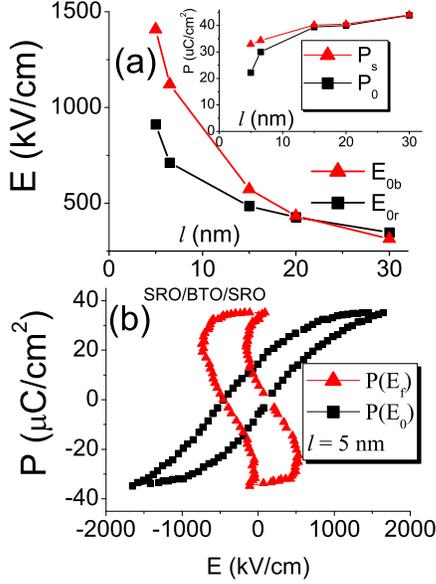}
\vskip -0.5mm
\caption{ (a) The external field $E_{0b}$ (where $E_{f}=0$) and $E_{0r}$
where the relaxation of polarization starts in 5~nm thick film\protect\cite%
{KimL05}. Inset: the spontaneous polarization $P_{s}$ and the extrapolated $%
P_{0}$\protect\cite{KimL05}. (b) The measured $P(E_{0})$ and the ``actual'' $%
P(E_{f})$ hysteresis loops. }
\label{fig:noh1}
\end{figure}

\begin{figure}[tbp]
\includegraphics{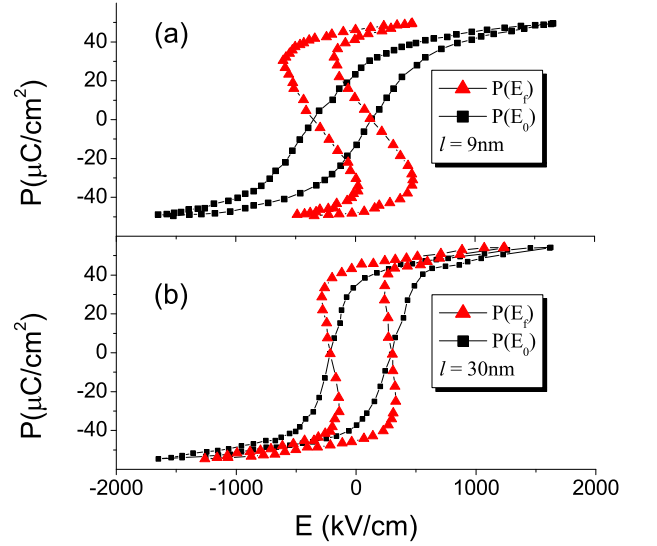}
\vskip -0.5mm
\caption{ The measured $P(E_{0})$ and the ``actual'' $P(E_{f})$ hysteresis
loops: (a) film thickness $l=9$~nm, (b) $l=$30~nm.}
\label{fig:noh2}
\end{figure}

One can see that the slope becomes positive at $y>3\psi /(1-\psi )$, yet
this result is beyond domain of applicability of the single-harmonic
approximation. In fact, we know that the slope is negative for the film with
domains at low temperatures, as has been predicted some time ago\cite%
{BLPRB01}: it is a hallmark of domain structure governed mainly by
electrostatics. Indeed, a net polarization of the domain structure is $\bar{P%
}=0$ in zero external field $E_{0}$. At small $E_{0}$ there will be positive
net polarization in external field because of growth of domains oriented
along the external field, and the resulting \emph{negative} field in the FE.
Thus, the \textquotedblleft dielectric function" of the film is negative, $%
\epsilon _{f}=1+4\pi d\bar{P}/dE_{f}|_{E_{f}=0}<0$ \cite{BLPRB01}. Comparing
theory with the data, the expression for the \textquotedblleft dielectric
constant", given by Eq.~(31) of Ref. \cite{BLPRB01}, can be simplified to
\begin{equation}
\epsilon _{f}\approx -\epsilon _{e}l/d.  \label{eq:negef}
\end{equation}%
Substituting the numbers for the BTO on SRO/STO\ film with thickness $l=5$%
nm, we find the theoretical value $\epsilon _{f}=-525$ for equilibrium
conditions while the experimental one found from Fig.~\ref{fig:noh1}1b (raw
data for $2$kHz, Ref.~\cite{KimAPL05}) is $\epsilon _{f}=-680$, i.e. both
estimated and measured values are pretty close. The slope is negative in all
samples with thicknesses up to 30nm studied in Ref.~\cite{KimL05}, see Fig.~%
\ref{fig:noh2}.


\section{Experimental example: BaTiO$_{3}$ films with SrRuO$_{3}$ electrodes
on SrTiO$_{3}$ substrate}

Our above assumption of a second order phase transition may be a reasonable
approximation for BaTiO$_{3}$ films on SrRuO$_{3}$/SrTiO$_{3}$ (BTO/STO)
substrate that have been investigated down to thickness $l=5$nm\cite%
{KimL05,KimAPL05,nohNodl06}. We use it to illustrate our theoretical results
while discussing, in particular, the effects of the higher order terms in
the LGD free energy. Because of $-2.2\%$ lattice mismatch, the BTO film
effectively becomes a uniaxial FE and can be treated within the
Landau-Ginzburg-Devonshire (LGD) free energy very accurately (results are
precisely the same as in ab-initio calculations where the latter are
available\cite{BLapl06}.)

Our aim is to describe what we think to be a consistent method of
interpreting the experimental data, i.e. the ferroelectric hysteresis loops
for the BaTiO$_{3}$ films on SrRuO$_{3}$/SrTiO$_{3}$ with different
thicknesses, from $5$ to $30$nm. The electrode parameters were found to be
\cite{KimL05}: $d/2=\lambda =0.8$\AA , $\epsilon _{e}=8.45$. The parameters
of the material, i.e. the LGD coefficients for the homogeneous states are
taken from \cite{LiCross05} with a renormalization due to the lattice misfit
strain and the clamping according to \cite{pertsev98}, see the Appendix A.
This set is different from the one used in Ref.\cite{pertsev98} and there
are qualitative differences in some experimentally verifiable conclusions
for the two sets. In particular, the coefficients of \cite{LiCross05} imply
a weak first order paraelectric-ferroelectric phase transition, while those
of \cite{pertsev98} imply a second order one, which is also suggested by the
results of the first-principles modeling\cite{ghosez03}. Since the
definitive experimental studies were not carried out, it is not clear which
set is better, and we have chosen to use the set of Ref.\cite{LiCross05},
when discussing homogeneously polarized state, but the assuming second order
phase transition while treating the domain structures.

What remains undefined are the additional boundary conditions (\textrm{ABC}%
)\ parameters, but we show below that they can be neglected in the case of
BTO on SRO/STO. This can be seen from the analysis of the\ experimental data
for single domain states. The role of the \textrm{ABC} for homogeneously
polarized states has been studied in many papers in the present geometry
when the polar axis is perpendicular to the film plane, starting with
Kretschmer and Binder \cite{Kretschmer}. Specific features of a
ferroelectric surface/interface have been taken fully into account only
recently\cite{BL05}. For a symmetric system (two identical surfaces) and
single domain state their effect is reduced to a renormalization of the
coefficient $A,$ similar to that of Ref.~\cite{Kretschmer} $A\rightarrow
A_{1}=A+\left( 2\alpha +\beta \right) /l$, where $\alpha $ and $\beta $ are
the parameters pertaining to a surface/interface. One can find the
polarization directly from the experiment which would correspond to a zero
external field, $E_{0}=0,$ if there were no domains. In Ref. \cite{KimL05}
it has been found by extrapolation of the high external field part the $%
p\left( E_{0}\right) $ curves to $E_{0}=0.$ Generalizing Eq.~(\ref{Pe}) to
account for the ABC and the higher order terms in the LGD free energy, we
obtain:
\begin{equation}
\left( A_{1}+\frac{4\pi d}{l\epsilon _{e}}\right)
p+Bp^{3}+Cp^{5}+Fp^{7}=E_{0},  \label{HoLGD}
\end{equation}%
i.e. for $E_{0}=0$
\begin{equation}
A_{1}+\frac{4\pi d}{l\epsilon _{e}}+Bp_{0}^{2}+Cp_{0}^{4}+Fp_{0}^{6}=0,
\label{A1}
\end{equation}%
and one can find $A_{1}\left( l\right) $ from the experimental data for $%
p_{0}\left( l\right) $. The equation for spontaneous polarization $P_{s}$
(polarization in the film with zero internal field, $E_{f}=0$):
\begin{equation}
A_{1}+Bp_{s}^{2}+Cp_{s}^{4}+Fp_{s}^{6}=0,  \label{Ps}
\end{equation}%
and find it as a function of $l$. Since we know from the data the
polarization at zero external field, $p_{0}=p(E_{0}=0)$\cite{KimL05}, we can
easily obtain $p_{s}$ from (\ref{A1}),(\ref{Ps}). The functions $%
p_{e0}\left( l\right) $ and $p_{s}\left( l\right) $ are shown in Fig.\ref%
{fig:noh1}(inset). We see that the dependence $p_{s}\left( l\right) $ is
weak, i.e. the ABC can be ignored, since the variation of the spontaneous
polarization $P_{s}$ with thickness due to the $(2\alpha +\beta )/l$ term is
small, and we shall assume $A_{1}=A.$

Let us now discuss what the theory tells about the state of films of
different thickness at $E_{0}=0$. Let us consider first the case of $B>0.$
This case has been discussed in Sec. IV. We shall start with this case and
then try to generalize it. From Eq.~(\ref{HoLGD}) for $E_{0}=0,$ one sees
that the solution $p=0$ (the paraelectric phase) becomes the only
possibility at $l<4\pi d/\left( -A\epsilon _{e}\right) .$ Using the above
value for the coefficient $A,$ we find that it happens at $l=l_{h}=3.5$nm.

However, we also need to study the stability of the paraelectric and
homogeneously polarized state with respect to the domain formation. First,
we calculate the value of critical dead layer $d_{m}$ for our system. The
value of the transversal dielectric constant, $\epsilon _{\perp },$ can be
calculated with the use of the LGD coefficients, and we find that $\epsilon
_{\perp }=218$ at room temperature (Appendix D). Then, using Eq.~(\ref{dm}),
we find that $d_{m}=0.34$\AA , i.e. in our case $d=5d_{m}.$ Using values of $%
\lambda ,\epsilon _{e}$ from Ref. \cite{KimL05}, calculating $\epsilon
_{\perp }$ using the coefficients of Ref. \cite{LiCross05} (Appendix D), and
the value of $\sqrt{D/4\pi }=0.2$\AA\ from Ref$.$\cite{shirane2} (Appendix
E), we obtain the following critical thickness for domains at room
temperature ($\epsilon _{\perp }=218)$:
\begin{equation}
l_{d}^{\text{RT}}=\left( \frac{\pi D}{\epsilon _{\perp }}\right)
^{1/2}/|A|\simeq 1.6\text{ nm.}
\end{equation}%
Since $d>d_{m}$, then $l_{d}^{\text{RT}}<l_{h}^{\text{RT}}$, and the phase
transition is into a multidomain state. The spatial distribution of a
spontaneous polarization is near sinusoidal at $l\gtrsim $ $l_{d}^{\text{RT}%
} $. Higher harmonics develop with increasing thickness, and the
polarization distribution tends to a conventional structure with narrow
domain walls. The half-period of the sinusoidal domain structure can be
estimated as $a_{c}=1.7 $ nm at the transition and as $a^{\text{RT}}=2.2$nm
for $l=5$nm from $a\approx \pi /k_{c}$ and (\ref{loss}).

It is instructive to consider the phase transition with regards to the
thickness at zero temperature, where we get $\epsilon _{\perp }=408,$ $%
l_{d}^{(0\text{K)}}=0.8\mathrm{nm}$, $l_{h}^{(0\text{K)}}=2.5$nm. The last
result (a homogeneous critical thickness of $2.5$nm) is remarkable, since it
practically\emph{\ coincides }with the ab-initio calculation for the
critical thickness of $2.4$nm in Ref.\cite{ghosez03}. The ground state of
the film is, however, not homogeneous but multidomain, and the domain
ferroelectricity appears in films thicker than $l_{d}^{(0\text{K)}}=0.8$nm,
which is the \emph{true \textquotedblleft critical size\textquotedblright }\
for ferroelectricity in FE films in the present study at zero temperature.

To use the results of Sec.~IV, it is convenient to express the variable $y$
as a function of the film thickness. Since $2Dk_{c}^{2}\propto l^{-1},$ one
can put $2Dk_{c}^{2}\left( l\right) =l_{c}2Dk_{c}^{2}\left( l_{c}\right)
/l=-Al_{c}/l.$ Then $\widetilde{A}\left( l\right) =A-Al_{c}/l=A\frac{l-l_{c}%
}{l}$. The parameter $\xi \left( l\right) =4\pi d/\left( \varepsilon
_{g}l+\epsilon _{b}d\right) -2Dk_{c}^{2}\left( l\right) $ can be obtained as
follows. The phase transition into the homogeneous state would occur at $%
l=l_{h}$, at $A+4\pi d/\left( \epsilon _{e}l_{h}+\epsilon _{b}d\right) =0$
(Sec.~IIB). Therefore, $A=-4\pi d/\epsilon _{e}l_{h}$, hence $4\pi d/\left(
\epsilon _{e}l+\epsilon _{b}d\right) \simeq $ $4\pi d/\epsilon
_{e}l=-Al_{h}/l$ and $\xi =-Al_{h}/l+Al_{c}/l=-A\left( l_{h}-l_{c}\right) /l$%
. Finally, we get $y=\left( l-l_{c}\right) /\left( l_{h}-l_{c}\right) $. We
see that the beginning of the memory for the case of a second order
transition from the paraelectric phase, which is at $y=3/2,$ corresponds to $%
l=4.5$nm. Recall that for this case the homogeneous phase becomes stable at $%
y>3$ (or for $l>7.3$nm) within our approximation, but this conclusion is
unreliable because, as we discussed above, the homogeneously polarized state
will almost certainly remain metastable.

Returning now to the experimental data for BTO on SRO/STO, we should mention
that the thinnest film ($5$nm) is thicker than the academic
\textquotedblleft limit for memory" even if $B>0$ and even more so for $B<0.$
This does not mean, however, that the sample has a real stable memory
because apart from the memory loss because of absolute instability of the
homogeneously polarized state there is also a memory loss because of domain
\emph{nucleation. }There is still no reliable theory of this process, but we
can gain some insight into the process by analyzing the experimental data.
No memory has been found with a lifetime longer than 1 second in BTO on
SRO/STO capacitors with thicknesses less than 30nm\cite{KimL05}. Kim \textit{%
et al} have found that during the observation time $t_{relax}=10^{3}$s the
domains begin to form only if the applied external field along the
polarization is less than $E_{0r}=910$kV/cm for 5\textrm{nm} sample (raw
data). We can immediately estimate from the results above that at $%
E_{0}=E_{0r}$ the domains begin to form only when there is a field in the
film $E_{f}=-\left( 490\pm 70\right) ~$kV/cm opposite to the polarization.
This is natural: it was known for a long time that in order to nucleate
domains one has to apply an \textquotedblleft activation field" $\left(
E_{a}\right) $\cite{Merz56}. Studying relatively thick BaTiO$_{3}$ samples
with $l=\left( 2.5-35\right) \times 10^{-3}$cm, Merz \cite{Merz56} found
that this field depends both on the waiting time $\tau $ and on the sample
thickness $l$ and have obtained an empirical formula for it:
\begin{equation}
E_{a}=\left( \alpha \ln \tau \right) /l.  \label{eq:Emerz}
\end{equation}%
In his thick samples, Merz identified the activation fields to be up to $20$%
~kV/cm, which looks quite high if one were to apply the Merz's scaling law (%
\ref{eq:Emerz})\ to Noh's samples \cite{KimL05}. Indeed, the thicknesses of
the films in Ref.~\cite{KimL05} are four orders of magnitude smaller, while
the fields are less than the two orders of magnitude larger. It is evident
that the Merz's formula cannot be applied literally to the films of all
thicknesses. (To be fair, Merz studied not the strained BaTiO$_{3},$ but
this hardly accounts for such a difference). However, within the interval of
thicknesses studied in Ref.~\cite{KimL05} the Merz's formula describes, at
least qualitatively, the dependence of the activation field on the film
thickness (see Fig.\ref{fig:noh1}). It is worth mentioning that at small
thicknesses the domain structure is nearly sinusoidal, and one can expect
weaker pinning compared to thicker films. It is hardly surprising that the
empirical Merz's formula obtained for conventional domain structure does not
apply to a sinusoidal one.

Using Eq.~(\ref{HoLGD}), one can find that the electric field in
short-circuited sample is about $1200$~kV/cm, exceeding the magnitude of the
estimated activation field. This means that in a short-circuited sample
single domain state relaxes faster (perhaps much faster)\ than in $10^{3}$s.
If the value of the activation field is defined by the thickness only and
not by properties of electrodes or an electrode-film interface, one can
speculate about the properties of electrodes that can facilitate a smaller
field in\ a short-circuited sample and a longer, at least $10^{3}$s,
retention of a SD state. We have found that for $l=5$nm such an electrode
should have $\lambda /\epsilon _{e}=d/2\epsilon _{e}<0.043$\AA . Indeed,
with such a hypothetical electrode the depolarizing field in an unbiased 5nm
sample would be less than 500 kV/cm discussed above. Since in the BTO
samples \cite{KimL05} the value $\lambda /\epsilon _{e}$ is about $0.1$\AA ,
it does not seem totally impossible to find such an electrode. This will
correspond to $d=0.73$\AA $\approx 2d_{m}$, the phase transition would be
still into a multidomain state. The homogeneously polarized state that is
metastable at $l_{h}=1.4$nm\ would be retained for at least $10^{3}$s in the
5nm BTO\ film with such an electrode according to the discussion above. One
can estimate the critical thicknesses for the domain states to see that here
too $l_{d}<l_{h}$\ (since $d>d_{m}),$ although the estimates should use more
accurate formula for $k_{c},$\ compared to (\ref{eq:kc2}), because $d$\ is
getting close to $d_{m}.$ While looking for possibilities for ferroelectric
memories, it seems more promising to optimize both the properties of the
electrode and the FE material. Recall that $d_{m}\propto D^{1/2}$ and the
constant $D$ proved to be particularly low for the (100) orientation of BaTiO%
$_{3}$ \cite{shirane2}. From a theory viewpoint, we see that one can play
with electrodes, materials, orientations of the film growth, and the misfit
strains to find a combination that may be suitable for the nanosize
ferroelectric memory elements.

\section{Conclusions}

We have described the application of the Landau theory to a problem of
(meta)stability of a homogeneous state in thin ferroelectric films. The
theory allows to better understand the data and gain more insight into the
practically important problem of memory (meta)stability in thin films. One
advantage is that the theory can start with relatively simple ground state.
We have proven, in particular, that switching of a polarization necessarily
goes via the intermediate domain state, the empirical fact well known from
experiment. The emerging domain structures in thin films are quite unusual,
the \textquotedblleft sinusoidal" ones. We show the distribution of the
polarization, which is reminiscent of the one for domains with abrupt
distribution of polarization (the Kittel structure\cite{kittel}.) To slow
down the polarization relaxation into a domain state, one can select
different electrodes or ferroelectrics, and the theory guides one as to what
parameters are favorable. For instance, ferroelectrics with large energy of
domain walls (large gradient, or inhomogeneous terms in the free energy)
should have better memory retention.

There are many open questions for both theory and experiment. For instance,
there is no acceptable theory of the polarization switching as of yet and,
therefore, theory cannot predict what parameters are needed to achieve a
needed memory retention. How the sinusoidal domains evolve with temperature
and/or thickness of the film into the standard domains with sharp domain
walls, remains to be investigated. Obviously, the continuous theory will
play very important role in this future work.

APL has been partially supported by Spanish MEC (MAT2006-07196) and CICYT
NAN-2004-09183-C10-07.

\appendix \label{historical}

\section{Historical note}

The mathematical equivalence of electrostatics and magnetostatics led, at
the beginning of development of \ the theory of ferroelectric domain
structure, to applying the relevant results of the theory of magnetic domain
structures to the case of ferroelectrics. The ferromagnetic domains were
postulated by Weiss in 1907. However, until 1935 their origin had not been
actually understood. Several authors tried to understand their origin within
statistical mechanics of an infinite medium. This was even the case with
such an outstanding scientist as F.Bloch,\cite{Bloch} curiously, in the same
paper where he considered the famous \textquotedblleft Bloch domain wall".
In 1935 Landau and Lifshitz \cite{LL35} pointed out futility of these
efforts and indicated the demagnetizing effect of surfaces as the origin of
ferromagnetic domains, i.e. the demagnetizing field existing in finite
magnetic bodies. Landau and Lifshitz proposed a fairly complete theory of
ferromagnetic domain structure and obtained some formulas which since then
were often attributed to other authors. For instance, the proportionality of
the width of the domains to the square root of a film thickness $l$ is
frequently attributed to Kittel (see also \cite{LL8}). Kittel highly
appreciated the Landau and Lifshitz work and has further developed their
theory in 1940's summarizing the Landau and Lifshitz's and his own results
in his often cited review \cite{kittel}. For later development of the
ferromagnetic domain theory see, e.g., a textbook \cite{Hubert}.

In the theoretical part of their pioneering paper on the domain structure of
Rochelle salt and KH$_{2}$PO$_{4},$ Mitsui and Furuichi \cite{mitsui}
pointed out the relevance of the earlier results for ferromagnets with
strong uniaxial anisotropy to their case. Also in 1950's the group of K\"{a}%
nzig \cite{Kanzig} clearly realized that the depolarizing field in
ferroelectrics should be much more important than in ferromagnets. They
argued that in a slab of an uniaxial ferroelectric, e.g. KH$_{2}$PO$_{4}$,
with the polar axis perpendicular to the slab plane no phase transition into
mono-domain state is possible and the only possibility of phase transition
into ferroelectric phase is with formation of a domain structure. This phase
transition should occur at a lower temperature and the lowering of
temperature is inversely proportional to the slab thickness. They tried to
estimate this lowering of the phase transition temperature to find what is
now called "the critical thickness for the ferroelectricity" in the case of
KH$_{2}$PO$_{4}$. They also clearly understood that their estimates are
irrelevant to the case of a three-axial ferroelectric such as BaTiO$_{3}$
where the closure domains (nowadays called \textquotedblleft vortices")
should form. The possibility of screening of the depolarizing field not only
by free charges from the electrodes but also by the charge carriers of the
material itself was also clearly realized at the very beginning of a
systematic study of ferroelectricity. One can see the relevant discussion,
in e.g. papers by the group of K\"{a}nzig \cite{Kanzig}.

Ivanchik \cite{Ivanchik} considered the possibility of a mono-domain state
in a slab of nonelectroded uniaxial ferroelectric taking into account the
screening of the depolarizing field by the carriers of the material, which
was treated as a semiconductor with a large band gap. Later he and others
\cite{Ivanchik et al.} took also into account the screening by metallic and
semiconductor electrodes. Also in 1960s, the effect of finite screening
length of metallic electrodes on dielectric measurements has been discussed
by several authors \cite{Mead61,Ku64,Simmons65}. Batra and Silverman \cite%
{BatraSil} criticized some details of an approach adopted by Ivanchik's
group and considered an effect of incomplete screening by the electrodes on
the phase transition into the ferroelectric state assuming that the phase
transition proceeds into a mono-domain state. This topic was later developed
in a series of papers by Batra \textit{et al}.\cite{Batra et al.}. Similar
to the Ivanchik's group, the main assumption of two series of paper,
stability of the mono-domain state was never checked although earlier
Bjorkstam and Oettel \cite{Bjorkstam} pointed out that incomplete screening
due to a near electrode dead layer may lead to the same domain structure as
in non-electroded samples. It is clear now that the assumption about a
monodomain state is almost never satisfied for realistic parameters of the
systems, at least close to the phase transition. Chensky \cite{Chensky}
substantially improved the K\"{a}nzig's group determination of lowering of a
phase transition temperature due to impossibility of transition into a
monodomain state in a nonelectroded slab of a ferroelectric with the polar
axis perpendicular to the slab.\textbf{\ }Selyuk considered screening of the
depolarizing field by both space and surface charges and found that below
the phase transition a monodomain state may become energetically more
favorable than the multi-domain one \cite{Selyuk}.

The next important step was the Chensky and Tarasenko paper mentioned in the
Introduction. Recently, the Chensky-Tarasenko approach to study different
cases of the domain structure formation at phase transitions has been used
by Bratkovsky and Levanyuk \cite{BLinh02} and by Stefanovich \textit{et al.}%
\cite{Stefanovich} for the case of phase transition in ferroelectric
periodic structures. Some other papers are cited in the main text. The
depolarization field screening due to charge carriers of the material and
those of the electrodes has been recently reconsidered by Watanabe \cite%
{Watanabe}. He obtained results substantially different from those of the
Ivanchik's group at the expense of making several important mistakes, he
also avoided any comparison of his results with those of the previous
authors.

This historical note does not pretend to represent an exhaustive review of
development of theory of the domain structures in ferroelectrics. We have
not mentioned many important works aiming at citing only the most important
ones falling into the scope of the present paper. In particular, we did not
mention many papers devoted to domain structures far from the phase
transition (one could call this case the "Kittel limit" versus the
"Chensky-Tarasenko limit") as well as papers related to ferro\emph{elastic}
domains. Although conceptually the latter topic is close to the theory of
180-degree domains formation considered in this paper, it is still well
beyond its scope.

\section{Additional boundary conditions}

The equation of state (\ref{eqstate1a}) relates electric field at a given
point not only to the polarization at the same point but also to its spatial
derivatives. In the electrodynamics of continuous media this is referred to
as an account for spatial dispersion of dielectric constant \cite{LL8}, \cite%
{Agranovich}. Differential equations of the electrodynamics with account for
the spatial dispersion are of higher order than in usual ("local")
electrodynamics, i.e. to specify among solutions of these equation those
that correspond to physical situation one needs more boundary conditions
than in the local electrodynamics. These are the \textquotedblleft
additional boundary conditions" (ABC). Their origin for different physical
systems is discussed in detail in \cite{Agranovich}.

Another way to ABC consists in applying the Landau theory of phase
transitions not to an infinite medium but to finite systems. For the first
time such an approach has been proposed in the paper by Ginzburg and Landau
\cite{GL50}, where the Landau theory has been used, in particular, to
analyze properties of thin superconducting films. The authors argued that at
the boundary superconductor-vacuum or superconductor-dielectric the
derivative of the order parameter along the normal to the surface should be
zero in the absence of a magnetic field. With such boundary conditions there
is no dependence of the phase transition temperature on the film thickness.
As far as we know, the first example of the phase transition temperature
dependent on the film thickness was given by Ginzburg and Pitaevsky \cite%
{GinzburgPit}, who considered superfluid phase transition in thin films of
liquid helium sandwiched between two solids. They argued that the order
parameter should be zero at liquid helium-solid interface. The influence of
this boundary condition can be easily understood considering the loss of
stability of the normal phase in the same way as in the present paper.
Replacing in Eq. (\ref{LGD}) $P_{z}$ by $\eta $ one obtains the free energy
of the Landau theory of phase transitions\cite{LLStatphys}. Since in the
case of superfluid transition there is no physical field conjugated to the
order parameter instead of Eq.~(\ref{eqstate1b}) one has%
\begin{equation}
A\eta -g\partial _{z}^{2}\eta =0,  \label{StabLandau}
\end{equation}%
with the boundary conditions $\eta \left( z=\pm l/2\right) =0$, where $l$ is
the film thickness. Since at $A>0$ the solutions of this equation are real
exponentials, no non-trivial solution satisfies the boundary conditions and
the normal phase is stable. At $A<0,$ it has a solution $\cos \left( \sqrt{%
-A/g}z\right) $ which satisfies the boundary conditions, if $\sqrt{-A/g}%
l=\pi $. This value of $A$ corresponds to the loss of stability of the
normal phase, i.e. the phase transition takes place at $A\left(
T=T_{cf}\right) =-\pi ^{2}g^{-1}l^{-2}.$ Within the Landau theory $%
A=A^{\prime }\left( T-T_{cb}\right) ,$ where $T_{cb}$ is the bulk transition
temperature. Therefore, $T_{cf}-T_{cb}\propto -l^{-2}.$

In 1960s, there was an intensive discussion of \ the "proximity effects",
i.e. the properties of a system consisting of a normal metal deposited on
top of a superconducting metal. For the superconductor film the main effect
was the lowering of the phase transition temperature. At the end, it has
been realized that for sufficiently thick layers of the two materials and in
the absence of magnetic field this effect can be described within the
Ginzburg-Landau theory, i.e. the Landau theory for superconductors completed
by the boundary conditions:%
\begin{equation}
\lambda \frac{d\eta }{dz}\pm \eta =0,  \label{ABC}
\end{equation}%
for $z=\pm l/2,$ where $l$ is the thickness of the superconductor film
sandwiched between layers of a normal metal and $\lambda $ is the parameter
of the interface, the so-called extrapolation length. Boundary conditions of
this form were proposed by de Gennes \cite{de Gennes} while considering
another system, the Josephson junction, while their general character was
understood by Zaitsev \cite{Zaitsev}. These de Gennes-Zaitsev boundary
conditions \cite{Deutscher} proved to be applicable well beyond the theory
of superconductivity and were derived later on for many different systems.

For $\lambda >0$, the only case considered by de Gennes and Zaitsev, the
phase transition in the film occurs again at a lower temperature than in the
bulk. Indeed, the \ solution of \ Eq.~(\ref{StabLandau}) is again $\cos
\left( \sqrt{-A/g}z\right) $ but to satisfy the boundary conditions the
value of $A$ should satisfy the equation%
\begin{equation*}
\tan \left( \kappa _{1}l/2\right) =\frac{1}{\lambda \kappa _{1}},
\end{equation*}%
where $\kappa _{1}=\sqrt{-A/g}.$ For $l\gg \lambda ,$ the solution of this
equation is $A\simeq -\pi ^{2}g^{-1}\left( l+\lambda \right) ^{-2}$ and,
therefore, the form of distribution of the order parameter at the phase
transition point is $\cos \left[ \pi z/\left( l+\lambda \right) \right] ,$
and this functional form explains the term "extrapolation length".

Later, these conditions appeared in study of ferromagnets derived from
purely phenomenological point of view by Kaganov and Omelyanchuk \cite%
{Kaganov}. These authors have taken into account dependence of the surface
energy of the crystal on the order parameter via terms $\alpha \eta
^{2}\left( \pm l/2\right) /2$ in the free energy per unit surface
considering ferromagnetic phase transition in a slab with the anisotropy
axis in the slab plane. In this geometry, no magnetic field arises due to
inhomogeneous magnetization. As a result, they have arrived at the condition%
\begin{equation}
g\frac{d\eta }{dz}\pm \alpha \eta =0,  \label{KO}
\end{equation}%
which is the same condition as Eq.~(\ref{ABC}). The most important novelty
of Ref.\cite{Kaganov} is that a possibility of the two signs of $\lambda
=g/\alpha $ ("positive and negative surface energy") has been discussed. In
the case of negative $\lambda ,$ a non-trivial solution is possible at $A>0$%
. It is $\cosh \left( \sqrt{A/g}z\right) ,$ which satisfies the boundary
conditions if%
\begin{equation*}
\tanh \left( \kappa _{2}l/2\right) =\frac{1}{\lambda \kappa _{2}},
\end{equation*}%
where $\kappa _{2}=\sqrt{A/g}$. For $l\gg \left\vert \lambda \right\vert ,$
this equation has the solution $\kappa _{2}=\left\vert \lambda \right\vert
^{-1},$ i.e. at $A=g\lambda ^{2}.$ We need to mention that now the phase
transition occurs at $A>0$ and the temperature of the transition is
independent of $l$, i.e. the same transition would occur at the surface of
an infinite half-space. The form of distribution of the order parameter at
the transition is characterized by exponential fall-off of the order
parameter while going from the surface into the bulk, i.e. the phase
transition is now not a bulk but a surface transition. The width of the
region affected by the transition increases approaching the temperature of
the transition in the bulk, so that all the results for the film are to be
affected by the transition. For an infinite half-space, this occurs at $A=0$
only. Of course, the Kaganov-Omelyanchuk treatment is valid also for
ferroelectrics with the polar axis lying in the slab plane as well as for
any order parameter which is not coupled with long-range fields.

Kretschmer and Binder \cite{Kretschmer}, while using the same boundary
conditions, have taken into account that for ferromagnets and ferroelectrics
with the magnetization (polarization) perpendicular to the surface the
surface effect will be affected by magnetic (electric) field arising due to
the inhomogeneities in the magnetization (polarization). For ferroelectrics,
their case is that of a slab with short-circuited ideal metallic electrodes,
i.e. of a transition into single-domain state. They found essential
differences with the case of\ absence of the depolarizing field, i.e. they
found essential differences between the cases of polarization parallel and
perpendicular to the surface. Tagantsev and Guerra \cite{Tagantsev} have
demonstrated recently that in this problem one needs to take into account
the noncritical part of the polarization. We shall follow them in this
regard.

Along with a modified Eq.(\ref{StabLandau}), which has now the form%
\begin{equation}
AP-g\partial _{z}^{2}P=E,  \label{StabKB}
\end{equation}%
one has to consider an electrostatics equation%
\begin{equation*}
\partial _{z}\left( \epsilon _{b}E+4\pi P\right) =0,
\end{equation*}%
or%
\begin{equation}
E+\frac{4\pi }{\epsilon _{b}}P=C=\frac{4\pi }{\epsilon _{b}l}%
\int_{-l/2}^{l/2}Pdz,  \label{electrostat}
\end{equation}%
where the constant $C$ is found from the condition of the short circuit: $%
\int_{-l/2}^{l/2}Edz=0$. Substituting Eq.(\ref{electrostat}) into (\ref%
{StabKB}) one obtains
\begin{equation*}
\left( A+4\pi /\epsilon _{b}\right) P-g\partial _{z}^{2}P=C,
\end{equation*}%
and has to look for a non-trivial solution of this equation satisfying the
boundary conditions (\ref{ABC}). The form of the solution does not depend
now on the sign of $A$, it is%
\begin{equation*}
P=\frac{C}{A+4\pi /\epsilon _{b}}+C_{1}\cosh \kappa z,
\end{equation*}%
where $\kappa =\sqrt{\left( A+4\pi /\epsilon _{b}\right) /g}$ and $C_{1}$ is
another constant related to $C$ via Eq.~(\ref{electrostat}). As a result,
one finds
\begin{equation*}
P=\frac{C}{A+4\pi /\epsilon _{b}}\left( 1+\frac{A\kappa l}{2\sinh \left(
\kappa l/2\right) }\cosh \kappa z\right) .
\end{equation*}%
The boundary conditions (\ref{ABC}) are satisfied if%
\begin{eqnarray}
A &=&-\frac{2}{\kappa l\left( \lambda \kappa +\coth \left( \kappa l/2\right)
\right) }\simeq -\frac{2}{\kappa l\left( \lambda \kappa +1\right) }  \notag
\\
&=&-\frac{2\alpha }{\kappa l\left( g_{1}\kappa +\alpha \right) }.
\label{TransKB}
\end{eqnarray}%
Since $\kappa ^{-1}$is the same order of magnitude as the interatomic
distances, the approximation in this equation is very good. The polarization
distribution at the transition is%
\begin{equation}
P\propto 1-\frac{\alpha }{\left( g\kappa +\alpha \right) \sinh \left( \kappa
l/2\right) }\cosh \kappa z.
\end{equation}%
One sees that now the phase transition occurs in the whole volume, both for
a "hard", $\alpha >0,$ and for a "soft", $\alpha <0$, with the polarization
that is practically constant over the volume being somewhat smaller for
"hard" surfaces and somewhat larger for the soft ones. The shift of the
phase transition temperature in the both cases is proportional to $l^{-1}$.

Since $\kappa ^{-1}$is very small, the formulas for the polarization
distribution near ferroelectric surfaces are beyond the continuous medium
theory used by Kretschmer and Binder and should be considered as
qualitative. In this sense account for the noncritical polarization is not
very essential here. Another shortcoming of their theory mentioned above is
overlooking the \textquotedblleft polar" character of a surface which is not
important for ferromagnets but is essential for ferroelectrics. This
importance has been emphasized by Levanyuk and Minyukov \cite{LevMin}. While
discussing surface structural phase transitions, these authors have pointed
out that when $\eta $ can be identified with polarization perpendicular to
the surface, the de Gennes-Zaitsev boundary conditions should be
complemented by an additional term which makes them inhomogeneous.
Bratkovsky and Levanyuk \cite{BL05} have explicitly taken into account the
complete boundary conditions considering phase transition in a slab of
uniaxial ferroelectric with ideal metallic electrodes. Because of the
inhomogeneity of the boundary conditions, the polarization is never zero
over all the volume, even in the paraelectric phase, though the region where
it is not zero is the same as the near-surface region in the Kretschmer and
Binder theory, i.e. its thickness is comparable to interatomic distances.
Still, to study stability of the paraelectric phase one has to linearize the
problem not with respect to $P=0,$ but with respect to $P_{e}\left( z\right)
,$ an inhomogeneous distribution existing in the paraelectric phase. As a
result, instead of Eq.(\ref{StabKB}) for $P$ one has an equation for $\delta
P=P-P_{e}\left( z\right) $ with the coefficient of the first term $%
A+BP_{e}^{2}\left( z\right) $ instead of $A$ in Eq.(\ref{StabKB}). This
means an additional "hardening" of the surface layer or an effective
positive renormalization of the coefficient $\alpha $ calculated in Ref.
\cite{BL05}. This means, e.g., that a surface which is "soft", $\alpha <0,$
for polarization parallel to the surface, may be "hard" with respect to
polarization perpendicular to the same surface. The polarity of surfaces
leads also to smearing of the phase transition if the two surfaces are
different \cite{BL05}.

Here, we have given only a very brief overview of the phenomenological
theory of the surface effect on ferroelectric phase transitions leaving
aside both microscopic theories and discussion of other effects of the
surfaces, e.g. their role in the polarization switching (see \cite%
{Tagantsev1}). In 1990s, when thin ferroelectric films became a popular
topic there appeared numerous papers reconsidering the phenomenological
theory and trying to find its parameters from the experimental data or
microscopic theories. We shall not review these papers because, to our
understanding, they did not advanced much the development of the
phenomenological theory in comparison with the papers cited above. This
certainly true of the formulation of the boundary conditions and finding the
phase transition temperature that we were concerned about in this Appendix.

We do not take into account the ABC in the present review. From the point of
view of electrodynamics of continuous media, this is allowed since the
"spatial dispersion of the dielectric constant" is taken into account in the
direction parallel to the surface ($x$) and, in effect, not in the direction
perpendicular to the surface ($z$): the coefficient $g$ does not enter our
final formulas. Physically, this is justified by the rigidity of
polarization with respect to its variation along $z-$axis, which we have
discussed while reviewing the results of Kretschmer and Binder. In addition,
the experimental data for the system (BTO\ on STO/SRO)\ we have used to
illustrate the theory indicates that the ABC can be neglected \cite{BLapl06}.

\section{Methodological note}

There is a lot of confusion in the literature with regards to the formula
for the energy of a ferroelectric with account for both the voltage source
and the electric field due to the polarization of the material. That is why
we preferred to use the equations of state as long as possible and, when
being forced to consider the free energy, we demonstrate its correctness by
comparing the results obtained from the equations of state and from
minimization of the free energy. By the way, the trouble with the Watanabe's
papers \cite{Watanabe} mentioned in the Historical note is due to his use of
the free energy while the Ivanchik's group used the equations of state to
treat the same problem.

To begin with, we note that some authors write the LGD functional as a
function of $\boldsymbol{P}$ while others did it as function of the
dielectric displacement $\boldsymbol{D}.$ The first group includes Ginzburg
\cite{Ginzburg}, Devonshire \cite{Devonshire}, Jona and Shirane \cite{Jona},
as well as many others, e.g. Batra \textit{et al}. and the present authors.%
{\normalsize \ }The second group includes importantly Lines and Glass \cite%
{Lines} and, e.g. Ivanchik \textit{et al.}\cite{Ivanchik}, \cite{Ivanchik et
al.}. There is also the third group led by Tagantsev \cite{Tagantsev86},
\cite{Tagantsev} who pointed out that in some situations it is important to
take into account that the order parameter does not refer to the total
polarization but only to a part of it and there is also another "background"
or "noncritical" contribution. In some problems it should be taken into
account when the effects of depolarizing field are considered. In this
review, unlike in our previous works, we have taken it into account but it
proved out that it disappears from the final results for the problems that
we considered here, which is not necessarily the case for some other problems%
{\normalsize . }

We think that the origin of existence of these variations in the functional
representations is the Landau and Lifshitz book "Electrodynamics of
Continuous Media" published in the USSR in 1957 and translated into English
in 1960 \cite{Landauelectr}. This excellent book had a mishap of using $%
\boldsymbol{D}$ instead of $\boldsymbol{P}$ exactly in the section devoted
to ferroelectrics. It has been eliminated in the second edition of the book
\cite{LL8},\ at least partially, but, because, perhaps, of a great authority
of Landau, the confusion originated.

The reason why Landau and Lifshitz put what we call now the LGD
thermodynamic potential as a function of $\boldsymbol{D}$ is that the
equilibrium thermodynamic properties of a system in a presence of an
external voltage source are described by thermodynamic potential depending
on $\boldsymbol{D}.$ This reasoning is flawed, however, because the
thermodynamic potential (or free energy) of the Landau theory of phase
transitions\cite{LLStatphys} is a \textit{non-equilibrium} thermodynamic
potential (free energy). It depends on the order parameter apart from the
conventional thermodynamic variables and only after minimization with
respect to the order parameter and substitution of its equilibrium value
into the original formula one obtains the equilibrium thermodynamic
potential. The order parameter describes the structural changes in the
material leading to the breaking of symmetry, which is the central idea of
this theory. In the case, of e.g. cubic-tetragonal phase transition in BaTiO$%
_{3}$ this structural change consists mainly in shift of the Ti ion with
respect to the Ba sublattice. Of course, the O ions also displace but to
change the symmetry from cubic to tetragonal the shift of Ti ions is
sufficient. Equally, a shift of only O ions without a shift of Ti ions would
lead to the same symmetry breaking. This illustrates the statement that in
the Landau theory the physical meaning of the order parameter does not
matter but only its symmetry properties. For ferroelectrics, this is
emphasized in the book by Strukov and Levanyuk \cite{Strukov} and is correct
while the effects of the{\normalsize \ }depolarizing field are not
considered. Quite often, it makes no difference whether to use $\boldsymbol{P%
}$ or $\boldsymbol{D}$ within the region of applicability of the "orthodox"
Landau theory, i.e. for small values of the order parameter and not far from
the phase transition. It is known, in particular, that in ferroelectrics
normally $\boldsymbol{P}\gg \boldsymbol{E}$, i.e.$4\pi \boldsymbol{P}\simeq
\boldsymbol{D}.$ However, even close to the phase transition one has to take
into account that the physical order parameter represents structural changes
with respect to which the system loses its stability at the transition,
while considering effects of the depolarizing field. In the case of proper
ferroelectrics, these structural changes can be related to a polarization
but not necessarily to the full polarization of the crystal. At the same
time, it is the latter which enters the electrostatics equations. This can
make it necessary to take into account the "background dielectric constant",
which is an independent phenomenological constant not entering the
"orthodox" Landau expansion, to be determined from the experimental data or
from the microscopic theory.

We see that much care is needed while applying the LGD theory to discuss
different phenomena in ferroelectrics. The examples where the proper care
was not taken are too numerous to be discussed here. One can say only that
the book by Strukov and Levanyuk \cite{Strukov} can be improved in this
regard. As an illustrative example, recall the boundary condition for the
order parameter imposed by Ivanchik \cite{Ivanchik} who considered a
homogeneously polarized ferroelectric slab with a depolarizing field
screened by the charge carriers of the material. The electrostatics dictates
that the value of $D_{n}$, should be zero at the surface. Since Ivanchik
considered $D_{n}$ as the order parameter entering the LGD functional he was
forced to conclude that the order parameter is zero at the surface. But what
does it mean? That the Ti or O ions in the surface layer are forced to stay
in the same positions as in the cubic phase? Nonsense. These ions can be
shifted, of course, but in such a way that at the surface $D_{n}=0.$ This is
clearly seen if one considers $P_{n}$ as the order parameter. One concludes
that at the surface $E_{n}=-4\pi P_{n}$ but these quantities are not
necessary zero and what are their values is a separate question. If one
takes into account the background dielectric constant ($\epsilon _{b}$) one
obtains{\normalsize \ }$E_{n}=-4\pi P_{n}/\epsilon _{b},${\normalsize \ }%
which does not change the qualitative result but may be important for
numerical estimations. Batra and Silverman \cite{BatraSil} (see Historical
note) were right in their criticism of Ivanchik's boundary condition in
spite of considering a case where this condition is not necessary and where
using either $P_{n}$ or $D_{n}$ as the order parameter makes no difference,
and it is also not necessary to take into account $\epsilon _{b}.$

There are several equivalent forms of writing down the free energy of
ferroelectric taking into account electric field created by the polarization
itself and by external voltage source in the cases where there is no other
source of the field rather than the ferroelectric polarization (like in a
non-electroded ferroelectric slab) or there is no electric field outside of
the ferroelectric and the voltage source (like in a ferroelectric slab with
the ideal metallic electrodes). We shall not discuss these forms here
because we are interested in a more general case. Let us mention only that a
careless generalization to this case of one of the above-mentioned formulas
leads quite often to errors, which could be avoided if the authors were to
compare at least some of their results with those obtained from the system
of equations of state and equations of electrostatics. Obviously, the
equations of state usually generate much less controversy than the forms of
the free energy.

In the present review, we use nearly the{\normalsize \ }same formula for the
free energy as Chensky and Tarasenko \cite{chensky82}, with the only
difference in writing a noncritical polarization term. They did not discuss
the origin of their formula but we shall explain why we think it is quite
natural. Conceptually, it follows the same line as, e.g. the calculation of
energy of collective excitations accompanied by long-range (electric or
magnetic) fields \cite{Agranovich}. First, one considers these excitations
without any long-range fields as if these fields were compensated by
external charges or currents, the so-called "mechanical excitons". Then, one
adds the energy of the long-range field created by the exciton without an
account for the changes, which such fields could produce in the exciton
itself, i.e. as if the fields were created in vacuum by the bound charges or
the Ampere currents of the "mechanical exciton". In our case, the
\textquotedblleft mechanical" part for the ferroelectric energy is given by
Eq.(\ref{LGD}), which does not contain any electric field and $P$ in this
formula has to be considered as a purely structural parameter. There is of
course also a "mechanical" noncritical part of the dead layer. The both have
the form of the first term{\normalsize \ }of the LGD formula (with another
coefficient, of course), because we do not take into account neither
nonlinearities (the second term) nor the spatial dispersion (the third and
the fourth terms) in the noncritical dielectric properties and in those of
the dead layer. It is shown in the text, see Eq.(\ref{elendiel}), that the
sum of the two energies, the "mechanical" and the "purely electric" ones,
gives what is used to be called the \textquotedblleft energy of electric
field in a dielectric". If we had considered explicitly a real metallic
electrode, which we avoided for simplicity sake, we would be obliged to take
into account not only the energy associated with the displacements of its
atoms and deformations of the filled electronic shells (the "background
dielectric constant"), but also an energy associated with redistribution of
electrons in the conduction band, while calculating the "mechanical"energy
of the electrode. This part is quite often overlooked but a detailed
discussion of this question would take too much space to do it here.

\section{Landau parameters for strained BaTiO$_{3}$}

The Landau parameters entering the equation of state of perovskite BaTiO$%
_{3} $ film on SrRuO$_{3}/$SrTiO$_{3}$ entering Eq.(\ref{eqstate1a})\ are
found from the parameters for the bulk BTO\ \cite{LiCross05} under
conditions of homogeneous in-plane misfit strain $%
u_{m}=(a-a_{0})/a_{0}=(b-b_{0})/b_{0}=-0.022$:
\begin{equation}
A=2a_{3}^{\ast },\quad B=4a_{33}^{\ast },\quad C=6a_{111},\quad F=8a_{1111},
\label{eq:par1}
\end{equation}%
where
\begin{equation}
a_{3}^{\ast }=a_{1}-u_{m}\frac{2Q_{12}}{s_{11}+s_{12}},\quad a_{33}^{\ast
}=a_{11}+\frac{Q_{12}^{2}}{s_{11}+s_{12}}.  \label{eq:par2}
\end{equation}%
The dielectric function in direction along the film $\epsilon _{\perp
}=1+\left( 2\epsilon _{0}a_{1}^{\ast }\right) ^{-1}$, where $a_{1}^{\ast
}=a_{1}-u_{m}\left( Q_{11}+Q_{12}\right) /\left( s_{11}+s_{12}\right) $ \cite%
{pertsev98}. When the strain drives transition second order from the weak
first order, one can approximate the dielectric function in a direction
along the easy axis $z$ and $P_{z}\equiv P_{3}$ (perpendicular to the film
plane) as $\epsilon _{c}=1+\left( 4\epsilon _{0}\left\vert a_{3}^{\ast
}\right\vert \right) ^{-1}$. The bulk BTO parameters appearing above are (in
SI\ units): $a_{1}=4.124\times 10^{5}(T-115)\quad $C$^{-2}$m$^{2}$N, where $%
T $ is the temperature in $%
{{}^\circ}%
$C, $a_{11}=-2.097\times 10^{8}\quad $C$^{-4}$m$^{6}$N$,$ $%
a_{111}=1.294\times 10^{9}\quad $C$^{-6}$m$^{10}$N$,$ $a_{1111}=3.863\times
10^{10}$C$^{-8}$m$^{14}$N, $Q_{11}=0.10,$ $Q_{12}=-0.034,$ $Q_{44}=0.029$ (C$%
^{-1}$m$^{2}$) \cite{yamada72}. The components of the compliance tensor for
BTO are $s_{11}=8.3\times 10^{-12},$ $s_{12}=-2.7\times 10^{-12},$ and $%
s_{44}=9.24\times 10^{-12}$ (in units of m$^{2}$N$^{-1}$).

Here in the text we are using the standard CGS units instead of the SI\
units, and the quantity we need to make estimates is the (dimensionless in
CGS)\ first Landau coefficient $A=\epsilon _{0}A_{\text{SI}}=-0.0054$, $%
\epsilon _{\perp }=218$ at room temperature. The coefficient $A_{\text{SI}}$
is found from (\ref{eq:par1}),(\ref{eq:par2}) and other relations given
above.

\section{Gradient term for BaTiO$_{3}$}

The gradient coefficient $D$ for BaTiO$_{3}$ was estimated from the neutron
scattering data \cite{shirane2} in the following way. The experimental data
gives the dependence of the soft mode frequency on the wave vector for the
directions perpendicular to (100). In terms of the present paper, the soft
mode frequency in the paraelectric phase can be obtained from the
polarization dynamic equation:
\begin{equation}
m\frac{d^{2}P_{\boldsymbol{k}}}{dt^{2}}+\left( A+Dk^{2}\right) P_{%
\boldsymbol{k}}=0,
\end{equation}%
where $m$ is a coefficient which will be determined, and $\boldsymbol{k}$ is
the wave vector in the plane perpendicular to (100). Then,
\begin{equation}
\omega ^{2}\left( k\right) =\frac{A+Dk^{2}}{m}.
\end{equation}%
We have found the coefficient $m$ from the value for $\omega \left( 0\right)
,$ and then estimated from the rest of the curve $\sqrt{\epsilon _{0}D_{%
\text{SI}}}=0.17$\AA\ ($D_{\text{SI}}$ is the gradient term in SI units, the
same estimate as we gave in Ref.\cite{BLapl06}). In the CGS units, used in
the present paper, the value is $\sqrt{D/4\pi }=0.17$\AA .

\end{document}